\documentclass[a4paper,11pt]{article}
\usepackage{mypackage}
\usepackage{physics}

\newcommand{\tpp}{{T_{++}}}

\title{On the positivity of light-ray operators}

\usepackage{authblk}
\author[1]{B. W. Freivogel}
\author[2]{H. Stoffels}
\affil[1]{ITFA and GRAPPA, University of Amsterdam, 904 Science Park, 1098 XH Amsterdam, the Netherlands}
\affil[2]{School of Mathematical and Computer Sciences, Heriot-Watt University, Edinburgh Campus, EH14 4AS Edinburgh, Scotland}

\date{\today}

\begin{document}

\maketitle

\begin{abstract}
    We consider light-ray operators $\L_{2n} = \int\d x^+ (x^+)^{2n}T_{++}$, where $x^+$ is a null coordinate and $n$ a positive integer, in QFT in Minkowski spacetime in arbitrary dimensions. These operators are generalizations of the average null energy operator, which is positive. We give a proof that the light-ray operators are positive in a non-minimally coupled but otherwise free scalar field theory, and we present various arguments that show that $\L_2$ is positive semi-definite in two-dimensional conformal field theories. However, we are also able to construct reasonable states which contradict these results by exploiting an infrared loophole in our proof. To resolve the resulting tension, we conjecture that the light-ray operators are positive in a more restrictive set of states. These states satisfy stronger conditions than the Hadamard condition, and have the interpretation of states that can be physically prepared. Our proposal is nontrivial even in two-dimensional CFT.
\end{abstract}

\section{Introduction}
All sensible classical theories obey the Null Energy Condition (NEC) \cite{classNEC_stab1,classNEC_stab2}, which states that for any null vector $l^a$, the stress-energy tensor $T_{ab}$ must obey $T_{ab}l^al^b\geq0$. The NEC plays a key role in constraining possible solutions of general relativity: exotic spacetimes such as traversable wormholes \cite{NECwormholes}, bouncing cosmologies \cite{BounceReview}, and black hole mimickers \cite{Penrose_1965} are all forbidden by the NEC. 

In contrast, every quantum field theory (QFT) allows for violations of the NEC \cite{EGJ_NietPuntsgewijs}, allowing \textit{some} of these exotic solutions to exist. For example, one can construct a traversable wormhole within the Standard Model of particle physics coupled to general relativity \cite{SM_Wormholes}. In order to understand what types of exotic solutions are possible, it is necessary to determine to what extent QFTs can violate the NEC.

In this paper, we are motivated by these gravitational questions but work in the simplest setting: a QFT in Minkowski spacetime, without gravity. Two important ways to implement the NEC in this context have been proposed. One is the Quantum Null Energy Condition (QNEC) \cite{QNEC,QNEC_proof}, which bounds $\smallmean{T_{ab}l^al^b}_\psi$ from below using the entanglement entropy. The other one is the Averaged Null Energy Condition (ANEC) \cite{ANEC_Entropy,ANEC_causaliteit}, which states that the NEC holds on average along a null geodesic with tangent vector $l^a$, i.e.
\begin{equation}
    \int_{-\infty}^\infty\mean{T_{++}}_\psi\d x^+\geq0\ ,
\end{equation}
where we assumed, without loss of generality, that $x^+ = x^0 + x^1$ is an affine parameter along the null geodesic, and $\ket{\psi}$ is any state of the field. Our goal is to determine whether any other energy conditions are generally true, and to that end we expand on our work in \cite{thesis} by considering the following light-ray operators (first introduced in \cite{LO_original} and further studied in e.g. \cite{LO_Mathys,LO_deBoer}):
\begin{align}
    \L_s = \int_{-\infty}^\infty (x^+)^s\,T_{++}\,\d x^+\ , \label{LO_def}
\end{align}
where $x^+$ is a null coordinate along the light-ray and we will further restrict to $s\in 2\mathbb{N}_0$. A particularly well-motivated constraint on these light-ray operators is suggested by the Conformally invariant Averaged Null Energy Condition (CANEC) \cite{CANEC_odd,CANEC_even}. When taken at face value and applied to a null geodesic embedded in the Minkowski spacetime lightcone of (without loss of generality) the origin, this condition would claim that
\begin{align}
    \int_{-\infty}^\infty\abs{x^+}^{d}\mean{T_{++}}_\psi dx^+\geq0\ ,\label{LO_NullConeCANEC}
\end{align}
for any state $\ket{\psi}$ and spacetime dimension $d$. For even $d$, \eqref{LO_NullConeCANEC} can be seen to constrain $\L_d$.

If these light-ray operators are positive\footnote{We often use simply `positive' to mean positive semi-definite.}, this would give an interesting set of additional constraints on negative null energy in QFT. We expect that this would be useful in the gravity applications mentioned above, among other reasons. In this paper, we present intriguing evidence for the positivity of these operators {\it in a particular class of states.}

We begin in section \ref{sec_FreeField} by presenting a proof from \cite{thesis} in free scalar field theory that many light-ray operators (dependent on the non-minimal coupling) are indeed positive; however, we immediately follow this with a calculation which finds counterexamples to the positivity of every $\L_{2n}$ except the ANEC operator $\L_0$. We repeat these steps for a conformal field theory (CFT) in section \ref{sec_CFT}, where we present various arguments for the positivity of $\L_2$ followed by an explicit counterexample. Finally, we discuss our results in more detail in section \ref{sec_DiscConc}.

Clearly, our counterexamples violate some assumptions in the proofs, so in section \ref{sec_DiscConc} we attempt to determine for what class of states the light-ray operators are positive. We establish the following data points:
\begin{itemize}
    \item States in free field theory which are superpositions of the vacuum and two particles with sufficiently weak support at vanishing null momentum ($p_+\rightarrow0$) satisfy positivity of $\L_{2n}$. To determine whether this support is sufficiently weak, considering $\L_{2(n+1)}$ might suffice: we suggest that $\mean{\L_{2n}}_\psi \geq 0$ if $|\smallmean{\L_{2(n+1)}}_\psi|<\infty$ and $\xi\leq(2n-1)/(8n)$, where $\xi$ is the non-minimal coupling.
    \item States in two-dimensional CFT prepared by acting with the stress tensor in a compact region of Euclidean time satisfy positivity of $\L_2$.
    \item States in two-dimensional CFT prepared by acting with sources in a compact region of Lorentzian spacetime have finite expectation values of $\L_{2n}$. If the relation between positivity of $\L_{2n}$ and finiteness of $\L_{2(n+1)}$ conjectured for the free field continues to hold for the two-dimensional CFT, then the finiteness of all $\L_{2n}$ implies their positivity.
\end{itemize}
Collecting these data points, we propose the following conjecture: {\it Light-ray operators are positive semi-definite in physically preparable states.} By `physically preparable states' we mean states that approach the vacuum sufficiently fast in the far IR and the far UV. 

Note that our proposal is nontrivial even in two-dimensional CFT, and therefore could be falsified or modified in the near future. In addition, the criterion of `physically preparable' states will need to be further developed and quantified.

\paragraph{Future directions and relation to other approaches.} Our results may be relevant to the programme initiated in \cite{HofmanMaldacena}, which studied constraints on the central charges of CFTs as they arise from the ANEC. If light-ray operators besides the ANEC operator are indeed positive, they could be used in a very similar way to further bound the central charges of CFTs. Furthermore, our results are  relevant to the energy flow operators defined in \cite{Zhiboedov} as
\begin{align}
    \E_\omega(n) = \frac{1}{n_\mu\bar{n}^\mu}\lim_{r\rightarrow\infty}r^{d-2}\int_{-\infty}^\infty\d t\,e^{-i\omega t (n_\nu\bar{n}^\nu)}T_{\rho\sigma}(rn + t\bar{n})\bar{n}^\rho\bar{n}^\sigma\ ,
\end{align}
where $n^\mu = (1,\ve{n})$ and $\bar{n}^\mu$ are a null vectors, with $n_\mu\bar{n}^\mu \neq0$. These operators measure the energy flux in a direction $\ve{n}$ reaching a detector at lightlike infinity if the temporal resolution of the detector is characterised by $\omega$ \cite{Zhiboedov}. Our light-ray operators can be realised as derivatives with respect to $\omega$ of $\E_\omega(n)$ at $\omega = 0$, and together form its real part. Furthermore, there may be a connection to the higher-spin ANEC \cite{ANEC_causaliteit,higher_spin_ANEC} (although we do not explore this connection in the present paper): in a free, minimally coupled scalar field, our light-ray operators are the momentum space analogues of the spinning operators constrained by the higher-spin ANEC. Finally, there is a connection between our work and \cite{local_NegativeEnergy}, which showed that regions with negative energy density cannot be localised arbitrarily far away from regions with positive energy density. Similarly, positivity of $\L_{2n}$ (for $n\geq1$) indicates that on a null geodesic, negative null energy densities must be flanked by positive null energy densities.

\section{Free field theory: proof and counterexamples}\label{sec_FreeField}

In this section, we start our investigation of the light-ray operators \eqref{LO_def} by considering them in a toy model: the non-minimally coupled but otherwise free real scalar field $\phi$ in $d$ spacetime dimensions. The classical action of this theory in a general background geometry is
\begin{align} 
    S = -\frac{1}{2}\int\d^dx\,\sqrt{-g}\roha{\nabla_a\phi\nabla^a\phi + (m^2 + \xi R)\phi^2}\ , \label{LO_NonMinAction}
\end{align}
where $m$ is the mass of the field, $\xi$ the non-minimal coupling to the Ricci scalar $R$, $g$ the determinant of the metric tensor and $\nabla_a$ the covariant derivative. A useful feature of this theory, besides its mathematical simplicity, is that $S$ is Weyl invariant if $m=0$ and $\xi = \frac{d-2}{4(d-1)}$, meaning that upon quantisation $\phi$ is either a simple QFT or a simple CFT, depending on the values of $m$ and $\xi$.

We first give a proof that there are many positive light-ray operators in this theory, and then present a class of states which provide explicit counterexamples to this positivity result.

\subsection{Proof of positivity in a free scalar field}\label{subsec_FreeProof}
The classical energy-momentum tensor associated with the action in \eqref{LO_NonMinAction} follows by varying $S$ with respect to the metric, leading to \cite{NonMinimalScalar}
\begin{align}
    T_{ab} = \nabla_a\phi\nabla_b\phi - \frac{1}{2} \roha{\nabla_c\phi\nabla^c\phi + m^2\phi^2}g_{ab} + \xi\roha{R_{ab} - \frac{1}{2}g_{ab}R + g_{ab}\nabla^c\nabla_c - \nabla_a\nabla_b}\phi^2\ . \label{LO_EMTgeneral}
\end{align}
The equations of motion for $\phi$ can be found by varying $S$ with respect to $\phi$, leading to
\begin{align}
    \roha{\nabla_a\nabla^a - m^2 - \xi R}\phi = 0\ . \label{LO_EoMscalar}
\end{align}
We now specialise to Minkowski spacetime, which has $R_{ab} = 0$, and choose (without loss of generality) the null geodesic to be the curve along which all spacetime coordinates vanish except $x^+ = x^0 + x^1$, which acts as the affine parameter. If $l^a$ is the tangent vector to this geodesic, then $l^a\nabla_a\phi = \diff{\phi}{x^+}$ and $T_{++} = T_{ab}l^al^b$ can be written as
\begin{align}
    T_{++} = T_{ab}l^al^b = \roha{\diff{\phi}{x^+}}^2 - \xi\diff{^2\phi^2}{x^{+\,2}}\ . \label{LO_NullNullEMT}
\end{align}
Furthermore, in Minkowski spacetime solutions to \eqref{LO_EoMscalar} can be decomposed in plane waves $\exp\roha{\pm ip_\mu x^\mu}$ with $p^2 = -m^2$. To canonically quantise the field, one defines the components of the four-momentum as $p^\mu = (\omega_{\ve{p}},\ve{p})$ such that $p_\mu x^\mu = -\omega_{\ve{p}}x^0 + \ve{p}\cdot\ve{x}$ and then promotes the coefficients in the plane wave decomposition of $\phi$ to creation and annihilation operators \cite{PeskinSchroeder}:
\begin{align}
    \hat{\phi}(x) = \int\frac{\d^{d-1}\ve{p}}{(2\pi)^{d-1}}\frac{1}{\sqrt{2\omega_\ve{p}}}\roha{\hat{a}_\ve{p}e^{ip\cdot x} + \hat{a}^\dagger_\ve{p} e^{-ip\cdot x}}\ ,
\end{align}
We are interested in $T_{ab}l^al^b$, which by inspection of \eqref{LO_NullNullEMT} means that the two-point function is an important quantity. Along the null geodesic, we have $p_\mu x^\mu = p_+x^+$ with $p_+ = -\frac{1}{2}(\omega_{\ve{p}} - p^1)$, and hence the (vacuum subtracted) two-point function on the null geodesic evaluates to
\begin{align}
    \normal{\hat{\phi}(x^+)\hat{\phi}(y^+)}\ \alis \hat{\phi}(x^+)\hat{\phi}(y^+) - \bra{0}\hat{\phi}(x^+)\hat{\phi}(y^+)\ket{0} \non
    \alis \int\frac{\d^{d-1}\ve{p}}{(2\pi)^{d-1}}\frac{\d^{d-1}\ve{k}}{(2\pi)^{d-1}}\frac{1}{\sqrt{4\omega_\ve{p}\omega_\ve{k}}}\roha{e^{-i(p_+ x^+ - k_+ y^+)}\hat{a}^\dagger_\ve{p}\hat{a}_\ve{k} + e^{i(p_+x^+ + k_+y^+)}\hat{a}_\ve{p}\hat{a}_\ve{k}} \non
    &\qquad\ \qquad\ \qquad\ \qquad\ \qquad\ \qquad\ \qquad\ \qquad\ \qquad\ \qquad\ \qquad\ \qquad\ + \mathrm{h.c.}\ , \label{LO_TwoPointGeneral}
\end{align}
where `h.c.' denotes the hermitian conjugate and $\ket{0}$ is the Minkowski vacuum. To simplify our subsequent calculations, we change integration variables from $p^1$ to $p_+$ using $\d p_+/p_+ = -\d p^1/\omega_{\ve{p}}$, leading to
\begin{align}
    \normal{\hat{\phi}(x^+)\hat{\phi}(y^+)}\ \alis \frac{1}{2}\int_{-\infty}^0\frac{\d p_+}{2\pi}\frac{\d k_+}{2\pi}\roha{\hat{A}^\dagger_p\hat{A}_ke^{-i(p_+x^+ - k_+y^+)} + \hat{A}_p\hat{A}_ke^{i(p_+x^+ + k_+y^+)}} + \mathrm{h.c.} \label{LO_TwoPointLightcone}
\end{align}
Here, $\ve{p}$ and $\ve{k}$ are functions of $p_+$ and $k_+$ respectively, and we define the shorthand $\hat{A}_p$ as
\begin{align}
    \hat{A}_p = \frac{1}{p_+}\int\frac{\d^{d-2}\ve{p}^\perp}{(2\pi)^{d-2}}\sqrt{\omega_\ve{p}}\,\hat{a}_\ve{p}\ . \label{LO_PencilOperator}
\end{align}
We construct $T_{++}$ from \eqref{LO_TwoPointLightcone} by applying the appropriate differential operator (as read off from \eqref{LO_NullNullEMT}) and taking the coincidence limit $y^+\rightarrow x^+$; this leads to
\begin{align}
    T_{++}(x^+) \alis \lim_{y^+\rightarrow x^+}\diff{}{x^+}\diff{}{y^+}\normal{\hat{\phi}(x^+)\hat{\phi}(y^+)} - \xi \diff{^2}{x^{+\, 2}}\normal{\hat{\phi}(x^+)\hat{\phi}(x^+)} \non
    \alis \frac{1}{2}\int_{-\infty}^0\frac{\d p_+}{2\pi}\frac{\d k_+}{2\pi}\left[\roha{p_+k_+ + (p_+ - k_+)^2\xi}e^{-i(p_+ - k_+)x^+}\hat{A}^\dagger_p\hat{A}_k \right. \non
    &\qquad\ \qquad\ \qquad\ \qquad\ \qquad \left. - \roha{p_+k_+ - (p_+ + k_+)^2\xi}e^{i(p_+ + k_+)x^+}\hat{A}_p\hat{A}_k \right] + \mathrm{h.c.}\label{LO_NullNullOperatorEMT}
\end{align}
We emphasise that this expression is only valid on this particular null geodesic.

Having constructed $T_{++}$ along the null geodesic, we turn to the light-ray operators. Upon substitution of \eqref{LO_NullNullOperatorEMT} into \eqref{LO_def} for $s = 2n$, we encounter $x^+$-integrals of the form
\begin{align}
    \int_{-\infty}^\infty\d x^+\,(x^+)^{2n}e^{-i\alpha x^+} \alis \frac{1}{(-i)^{2n}}\partial_\alpha^{2n}\int_{-\infty}^\infty\d x^+\,e^{-i\alpha x^+} = 2\pi(-1)^n\delta^{(2n)}(\alpha)\ . \label{LO_DeltaFunction}
\end{align}
The superscript on the delta function denotes the $2n^\text{th}$ derivative with respect to the argument of the delta function. Thus, substituting \eqref{LO_NullNullOperatorEMT} into \eqref{LO_def} for $s=2n$ and applying \eqref{LO_DeltaFunction} for $\alpha = \pm p_+ - k_+$ gives
\begin{align}
    \L_{2n} \alis \frac{(-1)^n}{4\pi}\int_{-\infty}^0\d p_+\d k_+\,\left[\roha{p_+k_+ + (p_+ - k_+)^2\xi}\hat{A}^\dagger_p\hat{A}_k\delta^{(2n)}(p_+ - k_+) \right. \non
    &\qquad\ \qquad\ \qquad\ \qquad\ \qquad\ \quad \left. - \roha{p_+k_+ - (p_+ + k_+)^2\xi}\hat{A}_p\hat{A}_k\delta^{(2n)}(p_+ + k_+) \right] + \mathrm{h.c.} \label{LO_Volledig}
\end{align}
Note that the argument of the second delta function, $p_+ + k_+$, only vanishes when $p_+ = k_+ = 0$ (since $p_+,k_+\leq0$ they cannot cancel each other). For a massive field, $p_+ = 0$ corresponds to an infinite momentum $p^1$, so if we restrict to states that don't excite modes of infinite momentum, expectation values of $\hat{A}_p\hat{A}_k$ in a massive theory are unsupported whenever $\delta^{(2n)}(p_+ + k_+)$ is supported. This allows us to ignore the contribution of the second delta function in \eqref{LO_Volledig} to any expectation value of $\L_{2n}$.

For $m=0$, the situation is more complex, although the same conclusion should hold. To see why, consider the expression for $p_+$ in terms of $\ve{p}$:
\begin{align}
    p_+ = -\frac{1}{2}\roha{\omega_\ve{p} - p^1} = -\frac{1}{2}\roha{\sqrt{m^2 + (\ve{p}^\perp)^2 + (p^1)^2} - p^1}\ .
\end{align}
It can be seen from this expression that $\ve{p}^\perp \neq \ve{0}^\perp$ fulfils the same role as a non-zero mass in ensuring that $p_+ < 0$ for finite $p^1$ (we still restrict to states which do not excite modes with infinite momentum). One may have $p_+ = 0$ at finite $p^1$ for $\ve{p}^\perp = \ve{0}^\perp$, but this is a single point in the $d-2$ dimensional space over which the integral in \eqref{LO_PencilOperator} is taken. Therefore, if we additionally restrict to states for which the expectation value of $\hat{a}_{\ve{p}}\hat{a}_{\ve{k}}$ is a sufficiently smooth and finite function of $\ve{p}$ and $\ve{k}$, then the contribution of $\ve{p}^\perp = \ve{0}^\perp$ to the integral in \eqref{LO_PencilOperator} vanishes. Thus, like for the massive theory, the second delta function in \eqref{LO_Volledig} does not contribute to expectation values of $\L_{2n}$ in a massless theory. Please note that this argument fails when $d=2$, since there is no $\ve{p}^\perp$ in that case.

Put together, we assume that we are working in $d\geq3$ or with a massive field, so that the second term in \eqref{LO_Volledig} can be ignored and we can focus on evaluating the contribution of the first term. This term naturally splits into a term which is independent of $\xi$ and one which isn't; we refer to these terms as the minimal and non-minimal coupling terms respectively. Let us consider them individually, and define $u = p_+ + k_+$ and $v = p_+ - k_+$; the minimal coupling term then becomes
\begin{align}
    \L_{2n}^\text{min} \alis \frac{(-1)^n}{8\pi}\int_{-\infty}^0\d u\int_{-u}^u\d v\, p_+k_+\hat{A}^\dagger_p\hat{A}_k\partial_v^{2n}\delta(v) + \mathrm{h.c.} \non
    \alis \frac{(-1)^{n+1}}{8\pi}\int_{-\infty}^0\d u \int_{-u}^{u}\d v\,\partial_v\roha{p_+k_+\hat{A}^\dagger_p\hat{A}_k}\partial_v^{2n-1}\delta(v) \non
    &\qquad\ \qquad\ \qquad\ \qquad\ \qquad\ \quad\ + \frac{(-1)^n}{8\pi}\int_{-\infty}^0\d u\, \viha{p_+k_+\hat{A}^\dagger_p\hat{A}_k\partial_v^{2n-1}\delta(v)}^u_{-u} + \mathrm{h.c.} \label{LO_FirstStepMin}
\end{align}
The step from the first to the second line is an integration by parts, for which we need to take the derivative of creation and annihilation operators. This is uncommon, but we interpret it as a formal operation which acquires its meaning inside an expectation value. To be explicit, consider the following state:
\begin{align}
    \ket{h} = \int h\roha{\ve{k}_1,\ldots,\ve{k}_N}\prod_{j\in\N}\roha{\frac{\d^{d-1}\ve{k}_j}{(2\pi)^{d-1}}\frac{\hat{a}^\dagger_{\ve{k}_j}}{\sqrt{2\omega_{\ve{k}_j}}}}\ket{0}\ , \label{LO_GeneralState}
\end{align}
with $\N = \cuha{1,\ldots,N}$, the set of positive integers up to and including a positive integer $N$, and $h$ some sufficiently smooth function which vanishes sufficiently fast at large momenta. This state includes the possibility of exciting a single mode multiple times, since we are allowed to make $h$ sharply peaked when $\ve{k}_i - \ve{k}_j$ approaches $\ve{0}$ for any pair $i,j$. Very general states can therefore be constructed as superpositions of the vacuum and states like $\ket{h}$, with different choices of $h$ (and accordingly different $N$). We then define $\partial_{p_+}^n\hat{A}_p$ as the operator which acts on $\ket{h}$ as
\begin{align}
    \partial_{p_+}^n\hat{A}_p\ket{h} = \int\frac{\d^{d-2}\ve{p}^\perp}{(2\pi)^{d-2}}\sum_{i=1}^N\int \partial_{p_+}^n\roha{\frac{h\roha{\ve{k}_1,\ldots,\ve{k}_i=\ve{p},\ldots,\ve{k}_N}}{\sqrt{2}\,p_+}}\prod_{j\in\N\setminus i}\roha{\frac{\d^{d-1}\ve{k}_j}{(2\pi)^{d-1}}\frac{\hat{a}^\dagger_{\ve{k}_j}}{\sqrt{2\omega_{\ve{k}_j}}}}\ket{0}\ .
\end{align}
The rationale behind this definition is that this is what one would get by acting on $\ket{h}$ with $\hat{A}_p$ and then differentiating $n$ times. With this construction, we have converted the derivatives of an operator to the familiar derivatives of functions, allowing us to define e.g.
\begin{align}
    \partial_{p_+}\hat{A}_{f(p_+)} \alis \roha{\partial_{p_+}f}\partial_f\hat{A}_{f(p_+)} \\
    \partial_{p_+}\roha{\alpha\hat{A}_p + \beta\hat{A}_{f(p_+)}} \alis \alpha \partial_{p_+}\hat{A}_p + \beta\partial_{p_+}\hat{A}_{f(p_+)} \\
    \partial_{p_+}\roha{\hat{A}_p\hat{A}_{f(p_+)}} \alis (\partial_{p_+}\hat{A}_p)\hat{A}_{f(p_+)} + \hat{A}_p(\partial_{p_+}\hat{A}_{f(p_+)})\\
    \partial_{p_+}\roha{f(p_+)\hat{A}_p} \alis (\partial_{p_+}f)\hat{A}_p + f\partial_{p_+}\hat{A}_p
\end{align}
for constant $\alpha,\beta\inC$ and a differentiable function $f$.

Having defined the derivative of an operator, we now consider the boundary terms in \eqref{LO_FirstStepMin}. They can be treated using the arguments we used to ignore the second delta function in \eqref{LO_Volledig}: the derivative of the delta function is only supported when its argument (here $\pm u$) vanishes, which does not happen at finite momenta for the massive field. For the massless field, a non-zero $\ve{p}^\perp$ takes on the role of the mass while the integral in \eqref{LO_PencilOperator} prevents the modes with $\ve{p}^\perp = \ve{0}^\perp$ and finite $p^1$ from contributing to the expectation value. Furthermore, we restrict to states in which modes of infinite momentum are not excited, so that the expectation value of $\hat{A}^\dagger_p\hat{A}_k$ vanishes sufficiently fast when $p_+ + k_+\rightarrow0$. Putting this together with our earlier assumption that either $m\neq0$ or $d\geq3$, we recognise that the boundary term in \eqref{LO_FirstStepMin} can be discarded. If we furthermore assume that the expectation value of any derivatives of $\hat{A}^\dagger_p\hat{A}_k$ also vanishes when $p_+,k_+\rightarrow0$, we can repeat this another $2n-1$ times, to obtain
\begin{align}
    \L_{2n}^\mathrm{min} \alis \frac{(-1)^n}{8\pi}\int_{-\infty}^0\d u\int_{-u}^u\d v\,\delta(v)\partial_v^{2n}\roha{p_+k_+\hat{A}^\dagger_p\hat{A}_k} + \mathrm{h.c.} \non
    \alis \frac{(-1)^n}{4\pi}\int_{-\infty}^0\d p_+\d k_+\,\frac{\delta(p_+ - k_+)}{2^{2n}}\sum_{j=0}^{2n}\roha{\begin{array}{c}
        2n \\
        j
    \end{array}}(-1)^{j}\partial_{p_+}^{2n-j}\partial_{k_+}^{j}\roha{p_+k_+\hat{A}^\dagger_p\hat{A}_k} + \mathrm{h.c.} \non
    \alis \frac{(-1)^n}{2^{2n+2}\pi}\sum_{j=0}^{2n}\roha{\begin{array}{c}
        2n \\
        j
    \end{array}}(-1)^j\int_{-\infty}^0\d p_+\,\partial_{p_+}^{2n-j}(p_+\hat{A}^\dagger_p)\partial_{p_+}^j(p_+\hat{A}_p) + \mathrm{h.c.}\ ,
\end{align}
where the step from the first to the second line used $\partial_v = \frac{1}{2}(\partial_{p_+} - \partial_{k_+})$. Since we have restricted to states and models in which modes with either $p_+ \rightarrow -\infty$ or $p_+ \rightarrow 0$ are sufficiently suppressed, we can integrate by parts in this expression while discarding the boundary terms until the $2n$ derivatives are distributed equally between $p_+\hat{A}^\dagger_p$ and $p_+\hat{A}_p$, leading to
\begin{align}
    \L^\mathrm{min}_{2n} \alis \frac{(-1)^n}{2^{2n+2}\pi}\sum_{j=0}^{2n}\roha{\begin{array}{c}
        2n \\
        j
    \end{array}}(-1)^n\int_{-\infty}^0\d p_+\,\partial_{p_+}^n(p_+\hat{A}^\dagger_p)\partial_{p_+}^n(p_+\hat{A}_p) + \mathrm{h.c.} \label{LO_sum}
\end{align}
The sum of binomial coefficients can be evaluated using the following identity:
\begin{align}
    \sum_{j=0}^{2n}\roha{\begin{array}{c}
        2n \\
        j
    \end{array}} = 2^{2n}\ . \label{LO_SumIdentity}
\end{align}
Expanding the derivatives in \eqref{LO_sum}, we therefore arrive at the following expression:
\begin{align}
    \L^\mathrm{min}_{2n} \alis \frac{1}{4\pi}\int_{-\infty}^0\d p_+\,\roha{p_+\partial_{p_+}^n\hat{A}^\dagger_p + n\partial_{p_+}^{n-1}\hat{A}^\dagger_p}\roha{p_+\partial_{p_+}^n\hat{A}_p + n\partial_{p_+}^{n-1}\hat{A}_p} + \mathrm{h.c.} \non
    \alis \frac{1}{4\pi}\int_{-\infty}^0\d p_+\,\viha{(p_+)^2\partial_{p_+}^n\hat{A}^\dagger_p\partial_{p_+}^n\hat{A}_p + n(n-1)\partial_{p_+}^{n-1}\hat{A}^\dagger_p\partial_{p_+}^{n-1}\hat{A}_p} + \mathrm{h.c.}\ , \label{LO_FinalMin}
\end{align}
where we combined the cross terms with the quadratic terms via integration by parts. Note that the derivatives in the second term of \eqref{LO_FinalMin} are not sensible for $n=0$; since the $n$-dependent prefactor of this term vanishes in this case, we don't consider this a problem.

Next, we consider the non-minimal coupling term. Using the same $u$ and $v$ as in \eqref{LO_FirstStepMin} and using the same arguments to discard the boundary terms that arise in the integration by parts, we find that we can express this term as
\begin{align}
    \L^\mathrm{non-min}_{2n} \alis \frac{(-1)^n\xi}{8\pi}\int_{-\infty}^0 \d u\int_{-u}^u\d v\,v^2\hat{A}^\dagger_p\hat{A}_k\partial_v^{2n}\delta(v) + \mathrm{h.c.} \non
    \alis \frac{(-1)^n2n(2n-1)\xi}{8\pi}\int_{-\infty}^0 \d u\int_{-u}^u\d v\,\delta(v)\partial_v^{2n-2}\roha{\hat{A}^\dagger_p\hat{A}_k} + \mathrm{h.c.}\ , \label{LO_FirstStepNonMin}
\end{align}
where we kept only those terms in the expansion of $\partial_v^{2n}(v^2\hat{A}^\dagger_p\hat{A}_k)$ that have precisely two derivatives acting on $v^2$; after all, more derivatives would annihilate $v^2$, while fewer would lead to surviving factors of $v$ that vanish upon evaluation of the integral over $v$. Note that for $n=0$, the derivatives in the final expression do not make sense; however, since the prefactor vanishes in this case, we do not consider that to be a problem.

By using that $\partial_v = \frac{1}{2}(\partial_{p_+} - \partial_{k_+})$, \eqref{LO_FirstStepNonMin} can be written in terms of $p_+$ and $k_+$ again, after which the delta function can be used to evaluate the integral over $k_+$:
\begin{align}
    \L^\mathrm{non-min}_{2n} \alis \frac{(-1)^n2n(2n-1)\xi}{2^{2n}\pi}\sum_{j=0}^{2n-2}\roha{\begin{array}{c}
        2n-2 \\
        j
    \end{array}}(-1)^j\int_{-\infty}^0 \d p_+ \,\partial_{p_+}^{2n-2-j}\hat{A}_p^\dagger\partial_{p_+}^j\hat{A}_p + \mathrm{h.c.}
\end{align}
For the non-minimal coupling term, the last step is to apply the procedure that led to \eqref{LO_sum}, i.e. integrate by parts to distribute the $2n-2$ derivatives equally between $\hat{A}^\dagger_p$ and $\hat{A}$, upon which we can apply the identity \eqref{LO_SumIdentity} to evaluate the sum over binomial coefficients. The outcome is
\begin{align}
    \L^\mathrm{non-min}_{2n} \alis -\frac{2n(2n-1)\xi}{4\pi}\int_{-\infty}^0\d p_+\,\partial_{p_+}^{n-1}\hat{A}^\dagger_p\partial_{p_+}^{n-1}\hat{A}_p + \mathrm{h.c.} \label{LO_FinalNonMin}
\end{align}
The overall minus sign arises due to the fact that we now have $2(n-1)$ derivatives to distribute, instead of the $2n$ derivatives in the minimal coupling term. Combining \eqref{LO_FinalMin} and \eqref{LO_FinalNonMin} we obtain the following expression for the light-ray operators:
\begin{align}
    \L_{2n} \alis \int_{-\infty}^0\frac{\d p_+}{2\pi}\viha{(p_+)^2\partial_{p_+}^n\hat{A}^\dagger_p\partial_{p_+}^n\hat{A}_p + n\roha{n-1-2\xi(2n-1)}\partial_{p_+}^{n-1}\hat{A}^\dagger_p\partial_{p_+}^{n-1}\hat{A}_p}\ .
\end{align}
For $n=0$, the second term in the square brackets vanishes because of the prefactor. We can then observe that the first term is a manifestly positive operator (multiplying an operator with its hermitian conjugate, so that any expectation value would be the norm of a state), meaning that $\L_0$ is positive. This is consistent with our expectations, since $\L_0$ is the ANEC operator.

To make further progress for $n\geq1$, we eliminate the explicit dependence on $p_+$ by defining
\begin{align}
    \hat{B}(p) = \sqrt{-\frac{p_+}{\mu}}\partial_{p_+}^{n-1}\hat{A}_p\ ,
\end{align}
with $\mu > 0$ some arbitrary fixed energy scale. If we then change variables to $q$ with $p_+ = -\mu e^{-q}$ and denote $\hat{B}_q = \hat{B}\roha{p(q)}$, we find that
\begin{align}
    \L_{2n} \alis \int_{-\infty}^\infty\frac{\mu e^{-q}\d q}{2\pi}\viha{\partial_q (e^{q/2}\hat{B}_q^\dagger)\partial_q (e^{q/2}\hat{B}_q) + n\roha{n-1-2\xi(2n-1)}e^q\hat{B}_q^\dagger\hat{B}_q} \non
    \alis \int_{-\infty}^\infty\frac{\mu\d q}{2\pi}\viha{\partial_q\hat{B}^\dagger_q\partial_q\hat{B}_q + \frac{1}{2}\partial_q(\hat{B}^\dagger_q\hat{B}_q) + \cuha{\frac{1}{4} + n(n - 1 - 2\xi(2n-1))}\hat{B}^\dagger_q\hat{B}_q}\ , \label{LO_FinalForm}
\end{align}
where we expanded $\partial_q (e^{q/2}\hat{B}_q^\dagger)\partial_q(e^{q/2}\hat{B}_q)$ and combined the cross terms into the full derivative $\partial_q(\hat{B}^\dagger_q\hat{B}_q)$; we can ignore this term since it is a boundary term which vanishes due to our assumption that no modes with infinite momentum have been excited. 

We finish our calculation by noting that the remaining operators in \eqref{LO_FinalForm} form a linear combination of operators multiplied by their respective hermitian conjugates, meaning that any expectation value would be a sum of (positive definite) norms of states. We can choose states in which the expectation value of $\hat{B}_q^\dagger\hat{B}_q$ is either oscillating rapidly or almost constant as a function of $q$, leading to the dominance of respectively the first or the last term in \eqref{LO_FinalForm}; hence, $\L_{2n}$ will be positive if and only if the term in curly brackets is non-negative. This condition is trivially satisfied for $n=0$, reproducing the ANEC \cite{Klinkhammer}; for $n\geq1$, we can formulate it in two equivalent ways, namely as a bound on $\xi$ given $n$ or as a bound on $n$ given $\xi$:
\begin{align}
    \xi \leq \frac{2n-1}{8n}& &2n\geq\frac{1}{1 - 4\xi}\ , \label{LO_NonMinConstraint}
\end{align}
where the form on the right assumes $\xi < \frac{1}{4}$; inspection of the form on the left reveals that there is no positive $\L_{2n}$ with $n\geq0$ if $\xi\geq\frac{1}{4}$.

Two values of $\xi$ are of particular interest: minimal coupling $\xi = 0$ and conformal coupling $\xi = \frac{d-2}{4(d-1)}$. For $\xi = 0$, \eqref{LO_NonMinConstraint} leads us to conclude that $\L_{2n}$ is positive for all integer $n\geq0$, whereas only those $\L_{2n}$ with $2n\geq d-1$ are positive for a conformally coupled theory. For the free scalar CFT in a spacetime with even $d$, we therefore see that $\L_d$, whose expectation values are constrained by the CANEC on the Minkowski lightcone, is indeed positive.

\subsection{Counterexamples in a free theory}\label{subsec_Counterexample}

In the previous section, a proof was given for the positivity of a range of light-ray operators. However, for this calculation to work, we had to restrict to a particular set of states. The central theme in this restriction is the behaviour of expectation values for large and small $p_+$: the calculation in section \ref{subsec_FreeProof} only holds if the expectation values of $\hat{A}_p\hat{A}_k$, $\hat{A}^\dagger_p\hat{A}_k$, and their derivatives vanish sufficiently quickly as $p_+, k_+ \rightarrow 0$ and $p_+,k_+\rightarrow-\infty$. To illustrate these restrictions more explicitly, we will construct a class of states in which $\L_{2n}$ is not positive. We do this by explicitly calculating the expectation value of $T_{++}$ and demonstrating that it can be made to have either sign by an appropriate choice of parameters.

\subsubsection{Choice of state and consistency conditions}

We consider states $\ket{\psi} = c_1\ket{0} + c_2\ket{2}$, where $c_1,c_2\inC$ with $\abs{c_1}^2 + \abs{c_2}^2 = 1$, and $\ket{2}$ is a normalised two-particle state constructed as
\begin{align}
    \ket{2} \alis \frac{1}{\sqrt{2}}\int\frac{\d^{d-1}\ve{p}}{(2\pi)^{d-1}}\frac{\d^{d-1}\ve{k}}{(2\pi)^{d-1}}\frac{p_+}{\sqrt{\omega_{\ve{p}}}}\frac{k_+}{\sqrt{\omega_{\ve{k}}}}f(\ve{p}^\perp)f(\ve{k}^{\perp})h(p_+,k_+)\hat{a}^\dagger_{\ve{p}}\hat{a}^\dagger_\ve{k}\ket{0}\ , \label{LO_Ket2Def}
\end{align}
where $f:\mathbb{R}^{d-2}\rightarrow\mathbb{C}$ and $h:\mathbb{R}^2\rightarrow\mathbb{C}$ are (sufficiently smooth) smearing functions, and we assume that $h(p_+,k_+) = h(k_+,p_+)$. The normalisation of $\ket{2}$ can be fixed by choosing $f$ and $h$ such that
\begin{align}
    \A \equiv \int\frac{\d^{d-2}\ve{p}^\perp}{(2\pi)^{d-2}}\abs{f(\ve{p}^\perp)}^2 = \roha{\int_{-\infty}^0\frac{\d p_+}{2\pi}\frac{\d k_+}{2\pi}p_+k_+\abs{h(p_+,k_+)}^2}^{-1/2}\ , \label{LO_Adef}
\end{align}
where we have defined the quantity $\A$ for future use. Similarly, we define \B\ as the integral of $f$:
\begin{align}
    \B = \int\frac{\d^{d-2}\ve{p}^\perp}{(2\pi)^{d-2}}\,f(\ve{p}^\perp)\ .
\end{align}
We guarantee that both \A\ and \B\ are finite by taking $f\in L^2$ to be Lebesgue integrable. 

To demonstrate that $\ket{\psi} = c_1\ket{0} + c_2\ket{2}$ allows for expectation values of $\L_{2n}$ with either sign, we first consider the expectation value of $T_{++}$ as written in \eqref{LO_NullNullOperatorEMT}. This operator naturally splits into a diagonal component (proportional to $\hat{a}^\dagger_{\ve{p}}\hat{a}_{\ve{k}}$ and its hermitian conjugate) and an off-diagonal component (proportional to $\hat{a}_{\ve{p}}\hat{a}_{\ve{k}}$ and its hermitian conjugate), which we denote by $T^\mathrm{diag}_{++}$ and $T^\mathrm{cross}_{++}$ respectively. Due to the vacuum subtraction, we find that
\begin{align}
    \smallmean{T^\mathrm{diag}_{++}}_{\psi} \alis \abs{c_2}^2\bra{2}T^\mathrm{diag}_{++}\ket{2} \non
    \alis -\abs{c_2}^2\A\abs{\B}^2\int_{-\infty}^0\frac{\d p_+}{2\pi}\frac{\d k_+}{2\pi}\frac{\d q_+}{2\pi}\roha{p_+k_+ + (p_+ - k_+)^2\xi}q_+\non
    &\qquad\ \qquad\ \qquad\ \qquad\ \qquad\ \qquad\ \quad \times h^*(p_+,q_+)h(q_+,k_+)e^{-i(p_+ - k_+)x^+} + \mathrm{c.c.}\ ,
\end{align}
where `c.c.'  denotes the complex conjugate. For the off-diagonal terms, we find instead that
\begin{align}
    \smallmean{T^\mathrm{cross}_{++}}_{\psi} \alis c_1^*c_2\bra{0}T^\mathrm{cross}_{++}\ket{2} + \mathrm{c.c.} \non
    \alis -\frac{c_1^*c_2\B^2}{\sqrt{2}}\int_{-\infty}^0\frac{\d p_+}{2\pi}\frac{\d k_+}{2\pi}\roha{p_+k_+ - (p_+ + k_+)^2\xi}h(p_+,k_+)e^{i(p_+ + k_+)x^+} + \mathrm{c.c.}
\end{align}
To make the number of minus signs more manifest, we change $p_+,k_+,q_+\rightarrow-p_+,-k_+,-q_+$ and define $h(p_+,k_+) = h(-p_+,-k_+)$, to obtain
\begin{align}
    \smallmean{T_{++}^\mathrm{diag}}_{\psi} \alis \abs{c_2}^2\A\abs{\B}^2\int_0^\infty\frac{\d p_+}{2\pi}\frac{\d k_+}{2\pi}\frac{\d q_+}{2\pi}\roha{p_+k_+ + (p_+ - k_+)^2\xi}q_+ \non
    &\qquad\ \qquad\ \qquad\ \qquad\ \qquad\ \qquad\ \quad\ \ \ \times h^*(p_+,q_+)h(q_+,k_+)e^{i(p_+ - k_+)x^+} + \mathrm{c.c.} \label{LO_2PemtDiag} \\
    \smallmean{T_{++}^\mathrm{cross}}_{\psi} \alis -\frac{c_1^*c_2\B^2}{\sqrt{2}}\int_0^\infty\frac{\d p_+}{2\pi}\frac{\d k_+}{2\pi}\roha{p_+k_+ - (p_+ + k_+)^2\xi}h(p_+,k_+)e^{-i(p_+ + k_+)x^+} + \mathrm{c.c.} \label{LO_2PemtCross}
\end{align}
At this point, let us introduce a specific choice of $h$ (recall that from now on, the light-ray momenta are taken to be non-negative):
\begin{align}
    h(p_+,k_+) = \frac{p_+^ak_+^a}{(p_+ + k_+)^b}e^{-l(p_+ + k_+)} \ , \label{LO_Counter_h}
\end{align}
where $a,b\geq0$ and $l\inR_{>0}$ is a length scale. We leave $f$ unspecified up to the condition \eqref{LO_Adef}; the possible values of $a$ and $b$ are restricted by various consistency conditions. Firstly, we demand that \A\ is both finite and non-zero, which constrains $a$ and $b$ based on \eqref{LO_Adef}. This can be seen by explicitly calculating the right-hand side of \eqref{LO_Adef}, for which we define the following function of three parameters $u,\,v$, and $\gamma$, assuming $u,v\geq0$ and $\Re(\gamma) > 0$:
\begin{align}
    I_{u,v}(\gamma) \alis \int_0^\infty\d x\,\d y\, \frac{x^uy^u}{(x+y)^v}e^{-(x+y)\gamma} \non
    \alis \frac{1}{\Gamma(v)}\int_0^\infty \d x\, \d y\, \d\alpha\,\alpha^{v-1}x^uy^ue^{-(x+y)(\gamma + \alpha)} \non
    \alis \int_0^\infty\frac{\d\alpha\,\alpha^{v-1}}{\Gamma(v)}\roha{\frac{\Gamma(u+1)}{(\gamma + \alpha)^{u+1}}}^2 \non
    \alis \frac{\Gamma^2(u+1)\Gamma(2u+2-v)}{\Gamma(2u+2)}\frac{1}{\gamma^{2u - v + 2}}\ . \label{LO_Iintegral}
\end{align}
In the first two steps, we made use of the following identity:
\begin{align}
    \frac{1}{x^z} = \frac{1}{\Gamma(z)}\int_0^\infty\d\alpha\,\alpha^{z-1}e^{-\alpha x}\ ; \label{LO_GammaIdentity}
\end{align}
the final step to \eqref{LO_Iintegral} was to apply the following definition of the Euler beta function:
\begin{align}
    B(z_1,z_2) = \int_0^\infty\frac{\alpha^{z_1 - 1}\d\alpha}{(1 + \alpha)^{z_1 + z_2}} = \frac{\Gamma(z_1)\Gamma(z_2)}{\Gamma(z_1 + z_2)}\ .
\end{align}
The integral representation is convergent for $\Re(z_1),\Re(z_2)>0$ or, in terms of the parameters present in \eqref{LO_Iintegral}, $v>0$ and $2u+2-v>0$. This condition is relevant because, according to \eqref{LO_Adef}, \eqref{LO_Counter_h} corresponds to a normalised state when $\A = 2\pi/\sqrt{I_{2a+1,2b}(2l)}$; this is finite since $a\geq0$ and $l > 0$, and as mentioned it is non-zero if $b>0$ and $2a + 2 - b > 0$.

The second consistency condition we impose is that $\ket{2}$ should be a Hadamard state, which means that the (non-vacuum-subtracted) two-point function $\bra{2}\hat{\phi}(x)\hat{\phi}(y)\ket{2}$ must have the divergence structure of $\bra{0}\hat{\phi}(x)\hat{\phi}(y)\ket{0}$ as $x\rightarrow y$ \cite{HadamardStates1,HadamardStates2}. To see whether this is the case, it suffices to check whether the normal-ordered two-point function $\bra{2}\normal{\hat{\phi}(x)\hat{\phi}(x)}\!\ket{2}$ is finite; if it is, then the divergence of $\bra{2}\hat{\phi}(x)\hat{\phi}(x)\ket{2}$ is the same as that in $\bra{0}\hat{\phi}(x)\hat{\phi}(x)\ket{0}$. Without restricting to points on the null geodesic, the expectation value of \eqref{LO_TwoPointGeneral} in the state $\ket{2}$ from \eqref{LO_Ket2Def} is given by
\begin{align}
    \bra{2}\!\normal{\hat{\phi}(x)\hat{\phi}(y)}\!\ket{2} \alis \int\frac{\d^{d-1}\ve{p}}{(2\pi)^{d-1}}\frac{\d^{d-1}\ve{k}}{(2\pi)^{d-1}}\frac{\d^{d-1}\ve{q}}{(2\pi)^{d-1}}\frac{p_+k_+q_+^2}{\omega_{\ve{p}}\omega_{\ve{k}}\omega_{\ve{q}}}f^*(\ve{p}^\perp)f(\ve{k}^\perp)\abs{f(\ve{q}^\perp)}^2 \non
    &\qquad \qquad \qquad \qquad \qquad \times h^*(p_+,q_+)h(q_+,k_+)e^{i(p\cdot x - k\cdot y)} + \mathrm{c.c.} 
\end{align}
Clearly, if we restrict to positive $f$ and take $h$ as in \eqref{LO_Counter_h}, the integrand consists of manifestly positive functions multiplied by a phase $e^{i(p\cdot x - k\cdot y)}$. Then, the expectation value achieves its maximum value when $x = y = 0$ (in this case, it is the integral of a positive function):
\begin{align}
    \bra{2}\!\normal{\hat{\phi}(0)\hat{\phi}(0)}\!\ket{2} \alis \int\frac{\d^{d-1}\ve{p}}{(2\pi)^{d-1}}\frac{\d^{d-1}\ve{k}}{(2\pi)^{d-1}}\frac{\d^{d-1}\ve{q}}{(2\pi)^{d-1}}\frac{p_+k_+q_+^2}{\omega_{\ve{p}}\omega_{\ve{k}}\omega_{\ve{q}}}f^*(\ve{p}^\perp)f(\ve{k}^\perp)\abs{f(\ve{q}^\perp)}^2 \non
    &\qquad \qquad \qquad \qquad \qquad \times h^*(p_+,q_+)h(q_+,k_+) + \mathrm{c.c.} \non
    \alis \A\abs{\B}^2\int_0^\infty\frac{\d p_+}{2\pi}\frac{\d k_+}{2\pi}\frac{\d q_+}{2\pi}h^*(p_+,q_+)h(q_+,k_+)q_+ + \mathrm{c.c.} \non
    \alis \A\abs{\B}^2\int_0^\infty\frac{\d p_+}{2\pi}\frac{\d k_+}{2\pi}\frac{\d q_+}{2\pi}\frac{p_+^ak_+^aq_+^{2a+1}e^{-l(p_+ + k_+ + 2q_+)}}{(p_+ + q_+)^b(q_+ + k_+)^b} + \mathrm{c.c.} \label{LO_NormalTwoPointEV}
\end{align}
The real exponential ensures that the integral does not diverge due to the behaviour of the integrand as $p_+\rightarrow\infty$; we can ensure that the integral does not diverge due to the small-$p_+$ behaviour of the integrand by demanding that $4a+4-2b > 0$, i.e. $2a+2-b>0$. Thus, if $\ket{2}$ is normalisable, then the divergent part of $\bra{2}\hat{\phi}(x)\hat{\phi}(x)\ket{2}$ is identical to that of $\bra{0}\hat{\phi}(x)\hat{\phi}(x)\ket{0}$, meaning that $\ket{2}$ is a Hadamard state.

Our third and final consistency condition is that $\ket{2}$ should have a finite energy. The classical expression for the Hamiltonian can be found using \eqref{LO_EMTgeneral}, namely as \cite{PeskinSchroeder}
\begin{align}
    H = \int\d^{d-1}\ve{x}\,T^{00} \alis \int\d^3\ve{x}\,\roha{\frac{1}{2}\dot{\phi}^2 + \partial_i\phi\partial^i\phi + m^2\phi^2 - \xi\partial_i\partial^i\phi^2}\ ,
\end{align}
where the dot denotes a derivative with respect to $x^0 = t$, $\partial_i$ with $i = 1,\ldots,d-1$ are the spatial derivatives, and we made use of the fact that we are working in Minkowski spacetime to set $R_{ab} = 0$ and to replace covariant derivatives by normal partial derivatives. Note that the final, $\xi$-dependent term only contributes a boundary term which we set to zero as usual. Hence, the classical Hamiltonian is the Hamiltonian of the minimally coupled free scalar field, regardless of $\xi$; quantising this expression and normal ordering it leads to the well-known result
\begin{align}
    \hat{H} = \int\frac{\d^{d-1}\ve{p}}{(2\pi)^{d-1}}\,\omega_\ve{p}\hat{a}^\dagger_\ve{p}\hat{a}_\ve{p} = \int\frac{\d^{d-2}\ve{p}^\perp}{(2\pi)^{d-2}}\int_0^\infty\frac{\d p_+}{2\pi}\frac{\omega_\ve{p}^2}{p_+}\,\hat{a}^\dagger_\ve{p}\hat{a}_\ve{p} \ .
\end{align}
The expectation value of $\hat{H}$ in $\ket{2}$ as defined by \eqref{LO_Ket2Def} is readily calculated to be
\begin{align}
    \bra{2}\hat{H}\ket{2} \alis 2\int\frac{\d^{d-2}\ve{p}^\perp}{(2\pi)^{d-2}}\frac{\d^{d-2}\ve{q}^\perp}{(2\pi)^{d-2}}\int_0^\infty\frac{\d p_+}{2\pi}\frac{\d q_+}{2\pi}\,q_+p_+\omega_{\ve{p}}\abs{f(\ve{p}^\perp)}^2\abs{f(\ve{q}^\perp)}^2\abs{h(p_+,q_+)}^2
\end{align}
The integral over $\ve{q}^\perp$ evaluates to \A\ by definition, but the evaluation of the integral over $\ve{p}^\perp$ is hindered by the presence of a factor of $\omega_{\ve{p}}$. However, from the definition $2p_+ = \omega_\ve{p} - p^1$ we can find that $\omega_\ve{p}p_+ = p_+^2 + \roha{(\ve{p}^\perp)^2 + m^2}/4$. Thus, the energy in $\ket{2}$ can be written as
\begin{align}
    \bra{2}\hat{H}\ket{2} \alis 2\A\left[\frac{1}{4}\int\frac{\d^{d-2}\ve{p}^\perp}{(2\pi)^{d-2}} \roha{(\ve{p}^\perp)^2 + m^2}\abs{f(\ve{p}^\perp)}^2\int_0^\infty\frac{\d p_+}{2\pi}\frac{\d q_+}{2\pi}\,q_+\abs{h(p_+,q_+)}^2 \right. \non
    &\qquad \qquad \qquad \qquad \qquad \qquad \qquad \qquad \left. +\ \A\int_0^\infty\frac{\d p_+}{2\pi}\frac{\d q_+}{2\pi}\,p_+^2q_+\abs{h(p_+,q_+)}^2 \right]\ . \label{LO_HamiltonianEV}
\end{align}
We do not choose a specific $f$, but we will restrict to $f$ for which the remaining integral over $\ve{p}^\perp$ is convergent. We can analyse the integrals over $p_+$ and $q_+$ by substituting \eqref{LO_Counter_h} and using the same arguments as we did to conclude that $\bra{0}\!\normal{\phi(0)\phi(0)}\!\ket{0}$ is finite: the real exponential ensures that the high-$p_+$ behaviour of the integrand does not lead to a divergence of the integral, while the small-$p_+$ behaviour of the integrand yields no divergences if $a$ and $b$ satisfy some conditions. For \eqref{LO_HamiltonianEV}, these conditions are $4a + 3 > 2b$ and $4a + 5 > 2b$ for the first and second set of integrals respectively. While the latter is implied by the earlier constraint $2a + 2 - b > 0$, the former is stricter. Thus, from now on we assume that $4a + 3 - 2b > 0$ as well as $a\geq0$ and $b > 0$ in \eqref{LO_Counter_h}; this ensures that $\ket{2}$ is normalisable and Hadamard, and that it has a finite energy.

\subsubsection{Computation of the counterexample}\label{subsubsec_CounterCalc}

Now that we have defined a particular class of states, we can compute the expectation value of $\L_{2n}$ explicitly; we start by calculating $\mean{T_{++}}_{\psi}$. Using \eqref{LO_Iintegral}, the off-diagonal contribution to $\mean{T_{++}}_{\psi}$ can be calculated quite straightforwardly. Substituting \eqref{LO_Counter_h} into \eqref{LO_2PemtCross}, we observe that
\begin{align}
    \smallmean{T_{++}^\mathrm{cross}}_{\psi} \alis -\frac{c_1^*c_2\B^2}{\sqrt{2}}\int_0^\infty\frac{\d p_+}{2\pi}\frac{\d k_+}{2\pi}\frac{\roha{p_+k_+ - (p_+ + k_+)^2\xi}p_+^ak_+^a}{(p_+ + k_+)^b}e^{-(p_+ + k_+)(l+ix^+)} + \mathrm{c.c.} \non
    \alis -\frac{c_1^*c_2\B^2}{4\pi^2\sqrt{2}}\roha{I_{a+1,b}(l+ix^+) - \xi I_{a,b-2}(l+ix^+} + \mathrm{c.c.} \non
    \alis -\frac{c_1^*c_2\B^2}{4\pi^2\sqrt{2}}\roha{\frac{a+1}{4a+6} - \xi}\frac{\Gamma^2(a+1)\Gamma(2a+4-b)}{\Gamma(2a+2)(l+ix^+)^{2a+4-b}} + \mathrm{c.c.}\ , \label{LO_Counter_Cross}
\end{align}
where the convergence of $I_{a+1,b}$ and $I_{a,b-2}$ is guaranteed by the condition $4a + 3 > 2b$. We continue by computing the contribution to $\mean{T_{++}}_{\psi}$ from diagonal terms. This starts similarly to the computation for the off-diagonal terms, namely by substituting \eqref{LO_Counter_h} into \eqref{LO_2PemtDiag}:
\begin{align}
    \smallmean{T_{++}^\mathrm{diag}}_{\psi} \alis \abs{c_2}^2\A\abs{\B}^2\int_0^\infty\frac{\d p_+}{2\pi}\frac{\d k_+}{2\pi}\frac{\d q_+}{2\pi}\frac{\roha{(1-2\xi)p_+k_+ + (p_+^2 + k_+^2)\xi}p_+^ak_+^aq_+^{2a+1}}{(p_+ + q_+)^b(q_+ + k_+)^b} \non
    &\qquad\ \qquad\ \qquad\ \qquad\ \qquad\ \qquad\ \times e^{-(l - ix^+)p_+}e^{-(l+ix^+)k_+}e^{-2lq_+} + \mathrm{c.c.} \non
    \alis \frac{\abs{c_2}^2\A\abs{\B}^2}{8\pi^3}\roha{(1-2\xi)J_{a+1,a+1}(l-ix^+) + \xi J_{a+2,a}(l-ix^+) + \xi J_{a,a+2}(l-ix^+)} \non
    &\qquad\ \qquad\ \qquad\ \qquad\ \qquad\ \qquad\ \qquad\ \qquad\ \qquad\ \qquad\ \qquad \qquad\ \qquad + \mathrm{c.c.}\ , \label{LO_DiagJ}
\end{align}
where we have defined the following function, again assuming $\Re(\gamma)>0$:
\begin{align}
    J_{p,q}(\gamma) \alis \int_0^\infty\d x\,\d y\,\d z\ \frac{x^py^qz^{2a+1}}{(x+z)^b(z+y)^b}e^{-\gamma x}e^{-\gamma^*y}e^{-(\gamma + \gamma^*)z} \non
    \alis \frac{1}{\Gamma^2(b)}\int_0^\infty\d x\,\d y\,\d z\,\d\alpha\,\d\beta\ \alpha^{b-1}\beta^{b-1}x^py^q z^{2a+1}e^{-(\gamma + \alpha)x}e^{-(\gamma^* + \beta)y}e^{-(\gamma^* + \alpha + \gamma + \beta)z} \non
    \alis \frac{1}{\Gamma^2(b)}\int_0^\infty\d\alpha\,\d\beta\, \frac{\alpha^{b-1}\Gamma(p+1)}{(\gamma + \alpha)^{p+1}} \frac{\beta^{b-1}\Gamma(q+1)}{(\gamma^* + \beta)^{q+1}}\frac{\Gamma(2a+2)}{(\gamma + \alpha + \gamma^* + \beta)^{2a+2}}\ , \label{LO_Jdef}
\end{align}
where we again made use of \eqref{LO_GammaIdentity}. The remaining integrals are difficult to compute, but since we eventually wish to integrate $\smallmean{T^\mathrm{diag}_{++}}_{\psi}$ over $x^+$ we are mainly interested in its behaviour when $x^+\rightarrow\pm\infty$. To find this behaviour, we first remark that for $4a+3>b$ and $\gamma = l - ix^+$ with $l > 0$, the relevant integrals converge. This can be seen by noting that the integrand has no singularities for finite $\alpha$ and $\beta$, and then changing to polar coordinates $r,\theta$ (with $r^2 = \alpha^2 + \beta^2$) to observe that the integrand (together with the Jacobian) decays as $1/r^{5-2b + p + q + 2a}$; since we are interested in cases where $p + q = 2a+2$, this is sufficiently fast to guarantee convergence.

We then consider $J_{p,q}(l-ix^+)$ for $x^+\rightarrow\infty$ (the case $x^+\rightarrow-\infty$ is completely analogous). To characterise the behaviour of $J_{p,q}$ in this limit, we make all quantities in the integral dimensionless using $x^+$ as a length scale: we define $\alpha' = \alpha/x^+$, $\beta' = \beta/x^+$, $\delta = l/x^+$, and a constant prefactor $\C_{p,q} = \Gamma(p+1)\Gamma(q+1)\Gamma(2a+2)/\roha{\Gamma^2(b)(-i)^pi^q}$ to obtain
\begin{align}
    J_{p,q}(l - ix^+) \alis \frac{\C_{p,q}}{(x^+)^{2a+4-2b+p+q}}\int_0^\infty\frac{\d\alpha'\d\beta'\,\alpha'^{b-1}\beta'^{b-1}}{(1 + i\delta + i\alpha')^{p+1}(1 - i\delta - i\beta')^{q+1}(2\delta + \alpha' + \beta')^{2a+2}} \ .
\end{align}
We now wish to isolate the contribution to the integral which dominates the limit $\delta\rightarrow0$. First, we can recognise that since $\delta\ll1$, we can Taylor expand $(1+i\delta+i\alpha')^{-p-1}$ and $(1 - i\delta-\beta')^{-q-1}$:
\begin{align}
    J_{p,q}(l-ix^+) \alis \frac{\C_{p,q}}{(x^+)^{2a+4-2b+p+q}}\viha{\int_0^\infty\! \frac{\d\alpha'\d\beta'\,\alpha'^{b-1}\beta'^{b-1}}{(1 + i\alpha')^{p+1}(1 - i\beta')^{q+1}(2\delta + \alpha' + \beta')^{2a+2}} + \order{\delta} } \non
    \equiv&\ \frac{\C_{p,q}}{(x^+)^{2a+4-2b+p+q}}\viha{K_{p,q}(\delta) + \order{\delta}}\ . \label{LO_Jscaling}
\end{align}
To find out how $K_{p,q}$ scales with $\delta$ for $\delta\ll1$, we choose some $\varepsilon$ with $2\delta\ll\varepsilon\ll1$ and split the integral $K_{p,q}$ in four parts, namely $0\leq\alpha',\beta'\leq\varepsilon$; $0\leq\alpha'\leq\varepsilon<\beta'$; $0\leq\beta'\leq\varepsilon<\alpha'$; and $\varepsilon<\alpha',\beta'$. In the latter three parts, we note that $\alpha'+\beta'\gg2\delta$, so that $(2\delta + \alpha' + \beta')^{-2a-2}$ can also be Taylor expanded, leaving a contribution that is independent of $\delta$ and $\order{\delta}$ terms. In the first part (with $0\leq\alpha',\beta'\leq\varepsilon$), we can Taylor expand $(1+i\alpha')^{-p-1}$ and $(1-i\beta')^{-q-1}$, yielding
\begin{align}
    K_{p,q}(\delta) \alis \int_0^\varepsilon\frac{\roha{1 + \order{\alpha'}}\roha{1 + \order{\beta'}}\alpha'^{b-1}\beta'^{b-1}\d\alpha'\d\beta'}{(2\delta + \alpha' + \beta')^{2a+2}} + \order{\delta^0}\ .
\end{align}
Note that factors of $\alpha'$ and $\beta'$ in the numerator effectively suppress the integral with factors of $\varepsilon$, so the dominant contribution is given by the term with as few of them as possible:
\begin{align}
    K_{p,q}(\delta) \alis \roha{1 + \order{\varepsilon}}\int_0^\varepsilon \frac{\alpha'^{b-1}\beta'^{b-1}\d\alpha'\d\beta'}{(2\delta + \alpha' + \beta')^{2a+2}} + \order{\delta^0}\ .
\end{align}
Let us consider positive integer values of $b$. We can then integrate by parts in the $\beta'$ integral:
\begin{align}
    K_{p,q}(\delta) \alis \roha{1 + \order{\varepsilon}}\int_0^\varepsilon\alpha'^{b-1}\left(\viha{-\frac{1}{2a+1}\frac{\beta'^{b-1}}{(2\delta + \alpha' + \beta')^{2a+1}}}_0^\varepsilon \right. \non
    &\qquad\ \qquad\ \qquad\ \qquad\ \qquad\ \quad\ \left.+ \frac{b-1}{2a+1}\int_0^\varepsilon\frac{\beta'^{b-2}\d\beta'}{(2\delta + \alpha' + \beta')^{2a+1}}\right)\d\alpha' + \order{\delta^0}\ .\label{LO_KintegralEstimate1}
\end{align}
Observe now that the boundary term at $\beta'=\varepsilon$ can once again be Taylor expanded in $\delta$, leading to a contribution which is $\order{\delta^0}$. The boundary term at $\beta'=0$ either vanishes (for $b > 1$) or it represents the relevant value of the integral (for $b=1$). In the latter case, we can immediately perform the $\alpha'$ integral as well:
\begin{align}
    K_{p,q}(\delta) \alis \frac{1 + \order{\varepsilon}}{2a+1}\int_0^\varepsilon\frac{\d\alpha'}{(2\delta + \alpha')^{2a+1}} + \order{\delta^0} = \frac{1 + \order{\varepsilon}}{(2a+1)2a}\frac{1}{(2\delta)^{2a}} + \order{\delta^0}\ ,
\end{align}
assuming that $a > 0$. In the case that $a = 0$ we have
\begin{align}
    K_{p,q}(\delta) = \roha{1+\order{\varepsilon}}\int_0^\varepsilon\frac{\d\alpha'}{2\delta + \alpha'} + \order{\delta^0} = -\roha{1 + \order{\varepsilon}}\ln\delta + \order{\delta^0}\ .
\end{align}
Plugging these expressions back into \eqref{LO_Jscaling} shows that for $b=1$, $a>0$, $J_{p,q}(l-ix^+)$ decays as $1/(x^+)^{p+q+2}$ when $x^+\rightarrow\infty$, while $J_{p,q}$ depends on $x^+$ as $\ln(x^+/l)/(x^+)^{p+q+2}$ in the same limit for $a=0$, $b=1$.

For the case of integer $b\geq2$, we return to \eqref{LO_KintegralEstimate1} and discard the boundary term at $\beta'=0$. This can be repeated $b-2$ times (which is possible since $4a + 3 > 2b$ implies $2a+2 > b$), to find
\begin{align}
    K_{p,q}(\delta) \alis \frac{(1 + \order{\varepsilon})\Gamma(b)\Gamma(b-1)}{\Gamma(2a+2)}\int_0^\varepsilon\frac{\alpha'^{b-1}\d\alpha'}{(2\delta + \alpha')^{2a+2-b}} + \order{\delta^0}\ .
\end{align}
We would like to perform the integral over $\alpha'$ as well, but there is no guaranteed ordering of the exponents $2a+2-b$ and $b-1$ while it is precisely this ordering which determines the relevant small $\alpha$-behaviour of the integrand. We therefore consider three cases separately: $2a + 2 > 2b$, $2a + 2 = 2b$, or $2a + 2 < 2b$.

If $2a + 2 > 2b$, we continue integrating by parts and discarding the boundary terms to find
\begin{align}
    K_{p,q}(\delta) = \frac{\Gamma^2(b)\Gamma^2(b-1)}{\Gamma(2a+2)\Gamma(2a+2-b)} \frac{1 + \order{\varepsilon}}{(2\delta)^{2a+2-2b}} + \order{\delta^0}\ .
\end{align}
When $2a+2 = 2b$, we can still integrate by parts $b-1$ times before anything is different:
\begin{align}
    K_{p,q}(\delta) = \frac{\roha{1 + \order{\varepsilon}}\Gamma(b)\Gamma(b-1)}{\Gamma(2b)}\int_0^\varepsilon\frac{\d\alpha'}{2\delta + \alpha'} = -\frac{\roha{1 + \order{\varepsilon}}\Gamma^2(b)}{(b-1)\Gamma(2b)}\ln\delta + \order{\delta^0}\ .
\end{align}
In the case that $2a+2 < 2b$, we observe that the integral is convergent in the limit $\delta \rightarrow 0$, meaning that the $\order{\delta^0}$-terms actually dominate the expansion in $\delta$. Thus, we conclude that for any value of $a\geq0$ and $b\in\mathbb{Z}_{\geq1}$, we have that asymptotically, $K_{p,q}$ scales as $\sim j_{a,b}(x^+)(x^+)^{2a+2-2b} + \order{(l/x^+)^0}$, with
\begin{align}
    j_{a,b}(x^+) = \begin{cases}
        1 & a > b-1 \\
        \ln(x^+/l) & a = b-1 \\
        0 & a < b-1
    \end{cases}\qquad .
\end{align}
If we plug this behaviour back into \eqref{LO_Jscaling}, we find that $J_{p,q}$ asymptotically depends on $x^+$ as
\begin{align}
    J_{p,q}(l-ix^+) \sim \frac{j_{a,b}(x^+)}{(x^+)^{p + q + 2}} + \frac{C}{(x^+)^{p+q + 2a + 4 - 2b}} = \frac{j_{a,b}(x^+)}{(x^+)^{2a + 4}} + \frac{C}{(x^+)^{4a + 6 - 2b}}\ , \label{LO_AsymptJ}
\end{align}
where $C$ is some $x^+$-independent number and the second expression takes into account the fact that we are interested in the case $p + q = 2a + 2$, according to \eqref{LO_DiagJ}.

We are now in a position to calculate $\mean{\L_{2n}}_{\psi} = \int\d x^+\,(x^+)^{2n} \smallmean{T_{++}^\mathrm{cross} + T_{++}^\mathrm{diag}}_{\psi}$. We begin with the cross term contribution, which using \eqref{LO_Counter_Cross} reads
\begin{align}
    \int_\mathbb{R}\d x^+\,(x^+)^{2n}\smallmean{T_{++}^\mathrm{cross}}_{\psi} = \D_{a,b,\xi}\int_{-\infty}^\infty\d x^+\roha{\frac{c_1^*c_2\B^2(x^+)^{2n}}{(l+ix^+)^{2a+4-b}} + \frac{c_1c_2^*(\B^*)^2(x^+)^{2n}}{(l-ix^+)^{2a+4-b}}}\ , \label{LO_LOCross}
\end{align}
where we have denoted all real numerical prefactors in \eqref{LO_Counter_Cross} by $\D_{a,b.\xi}$. \eqref{LO_LOCross} can be evaluated using the residue theorem if we choose $a$ and $b$ such that $2a+4-b = 2n+1$ (so that we have a first order pole) and $c_1$, $c_2$, and $f$ such that $c_1^*c_2\B^2\inR$ (so that the integrand in \eqref{LO_LOCross} vanishes sufficiently quickly as $x^+\rightarrow\infty$). Importantly, for this choice of parameters we find that the residue at either pole is finite and non-zero, meaning that the cross term contribution to $\mean{\L_{2n}}_{\psi}$ is finite and non-zero (up to the vanishing of $\D_{a,b,\xi}$, to which we return momentarily). Also note that the choice $2n + 1 = 2a+4-b$ is only possible for $n>\frac{3}{4}$.

For the contribution of the diagonal terms, we have not found a closed-form expression, but we have argued that $\smallmean{T_{++}^\mathrm{diag}}_{\psi}$ is non-singular for $x^+\inR$, and \eqref{LO_AsymptJ} gives the asymptotic behaviour. In particular, for $2a+4-b = 2n+1$, $(x^+)^{2n}\smallmean{T_{++}^\mathrm{diag}}_{\psi}$ asymptotically scales as
\begin{align}
    (x^+)^{2n}\smallmean{T_{++}^\mathrm{diag}}_{\psi} \sim \frac{j_{a,b}(x^+)}{(x^+)^{b+1}} + \frac{C}{(x^+)^{2n}}\ .
\end{align}
This decay is sufficiently fast to guarantee that $\int\d x^+\,(x^+)^{2n}\smallmean{T_{++}^\mathrm{diag}}_{\psi}$ converges for e.g. $a = n-1$ and $b = 1$ if $n\geq1$.

Thus, that there are choices of the parameters $a$, $b$, $n$, $c_1$, and $c_2$ for which both $\smallmean{T_{++}^\mathrm{cross}}_{\psi}$ and $\smallmean{T_{++}^\mathrm{diag}}_{\psi}$ make finite and non-zero contributions to $\mean{\L_{2n}}_{\psi}$. However, observe from \eqref{LO_Counter_Cross} and \eqref{LO_DiagJ} that the former is linear in $c_2$, while the latter is proportional to $\abs{c_2}^2$. Thus, by taking $\abs{c_2}\ll1$, we can make any non-zero contribution from $\smallmean{T_{++}^\mathrm{cross}}_{\psi}$ dominate the value of $\mean{\L_{2n}}_{\psi}$, and by choosing an appropriate sign for $c_2$ we can make this value negative, providing an explicit counterexample to the conclusion of section \ref{subsec_FreeProof}.

A potential issue with this counterexample is that the non-minimal coupling $\xi$ may be such that the factor $\D_{a,b,\xi}$ vanishes in \eqref{LO_LOCross}. From \eqref{LO_Counter_Cross}, we see that this happens whenever
\begin{align}
    \xi = \frac{a+1}{4a+6}\ .
\end{align}
This form suggests a method to circumvent this issue, as $\xi$ is a parameter of the theory, but $a$ is a parameter of the state under consideration. Thus, having chosen $a = n-1$ for $n\geq1$, if we do encounter a theory with $\xi = n/(4n+2)$ we simply choose a different state (e.g. with $a = n-\frac{1}{2}$ and $b=2$ for $n\geq2$, or $a = \frac{1}{2}$, $b=2$ for $n=1$) for which $\D_{a,b,\xi}$ does not vanish.

We therefore have counterexamples to the positivity of all $\L_{2n}$ with $n\geq1$. The states $\ket{2}$ from \eqref{LO_Ket2Def} with \eqref{LO_Counter_h} do not provide a counterexample for $n=0$, since the requirement that they should have a finite energy imposes the constraint $2n = 2a + 3 - b > \frac{3}{2}$. This is to be expected, since the free scalar field obeys the ANEC \cite{Klinkhammer}.

Finally, we consider how the counterexample presented in this section relates to the proof given in section \ref{subsec_FreeProof}. We do this by considering the expectation value of \eqref{LO_Volledig} in a state $\ket{\psi} = c_1\ket{0} + c_2\ket{2}$ with \eqref{LO_Ket2Def}, \eqref{LO_Counter_h}, and $2n+3=2a+4-b$. We focus in particular on the off-diagonal part, proportional to $\hat{A}_p\hat{A}_k$ and its hermitian conjugate; evaluating $\smallmean{\hat{A}_p\hat{A}_k}_{\psi}$ leads to
\begin{align}
    \smallmean{\L^\mathrm{cross}_{2n}}_{\psi} \alis \frac{(-1)^{n+1}c_1^*c_2\B^2}{2\pi\sqrt{2}}\!\int_0^\infty\!\d p_+\d k_+\!\viha{\frac{p_+^{a+1}k_+^{a+1}}{(p_+ + k_+)^b} - \frac{\xi p_+^{a}k_+^{a}}{(p_+ + k_+)^{b-2}}}\!e^{-l(p_+ + k_+)} \delta^{(2n)}(p_+ + k_+)\non
    &\qquad\ \qquad\ \qquad\ \qquad\ \qquad\ \qquad\ \qquad\ \qquad\ \qquad\ \qquad\ \qquad\ \qquad\ \ + \mathrm{c.c.}
\end{align}
To evaluate these integrals, we define $\bar{p} = p_+ + k_+$ and $\delta p = p_+ - k_+$, so that $p_+k_+ = \frac{1}{4}(\bar{p}^2 - \delta p^2)$ and hence
\begin{align}
    \smallmean{\L_{2n}^\mathrm{cross}}_{\psi} \alis -\frac{(-1)^nc_1^*c_2\B^2}{4^{a+1}\pi\sqrt{2}}\! \int_0^\infty\!\d\bar{p}\int_{-\bar{p}}^{\bar{p}}\!\d\delta p\!\viha{\frac{(\bar{p}^2 - \delta p^2)^{a+1}}{4\bar{p}^b} - \frac{\xi(\bar{p}^2 - \delta p^2)^{a}}{\bar{p}^{b-2}}}\!e^{-l\bar{p}}\delta^{(2n)}(\bar{p}) + \mathrm{c.c.} \non
    \alis -\frac{(-1)^nc_1^*c_2\B^2}{4^{a+2}\pi\sqrt{2}}\!\int_0^\infty\!\d\bar{p}\viha{\int_{-1}^1\!\d u\,(1-u^2 - 4\xi)(1-u^2)^a}\!\bar{p}^{2a+3-b}e^{-l\bar{p}}\delta^{(2n)}(\bar{p}) + \mathrm{c.c.} \non
    \alis -\frac{(-1)^nc_1^*c_2\B^2\F}{4^{a+2}\pi\sqrt{2}}\viha{\partial_{\bar{p}}^{2n}\roha{\bar{p}^{2a+3-b}e^{-l\bar{p}}}}_{\bar{p} = 0} + \mathrm{c.c.}\ , \label{LO_CrosstermDerivative}
\end{align}
where we defined $\F$ in the final line as the integral in square brackets in the second line, and the integral over $\bar{p}$ could be performed using integration by parts while ignoring boundary terms because the polynomial factor ensured that $\bar{p} = 0$ does not contribute and the exponential term suppressed any contribution from $\bar{p}\rightarrow\infty$. We now recall that we chose $2a+3-b = 2n$, meaning that there is one term from the remaining derivative which survives at $\bar{p} = 0$:
\begin{align}
    \smallmean{\L^\mathrm{cross}_{2n}}_{\psi} \alis -\frac{(-1)^nc_1^*c_2\B^2\F}{4^{a+2}\pi\sqrt{2}}\Gamma(2a+4-b) + \mathrm{c.c.}\ .
\end{align}
Clearly, this does not need to vanish, and it is now clear that this is because $h$ does not vanish sufficiently fast as $p_+,k_+\rightarrow0$ (note that $2a+3-b > 2n$, corresponding to a more rapidly decaying $h$, would suffice to obtain $\smallmean{\L^\mathrm{cross}_{2n}}_{\psi} = 0$).

\section{CFT examples and counterexamples}\label{sec_CFT}
We would like to know whether the above results are an artifact of working in free theories, or if they are generic. To address this, we work in another framework where things are somewhat tractable: CFTs. 

For simplicity, we will work in two-dimensional CFT; we follow the conventions of \cite{diFrancesco}. In this context, it is useful to write the light-ray operators $\L_{2n}$ from \eqref{LO_def} in terms of the Virasoro operators $L_n$. To this end, we first Wick rotate by setting $x^2 = ix^0$, and then define complex coordinates $z = ix^1 + x^2$ and $\bar{z} = -ix^1 + x^2$ so that $x^+ = -iz$ and $x^- = -i\bar{z}$; as usual, tracelessness of the energy-momentum tensor implies that $T_{z\bar{z}} = T_{\bar{z}z} = 0$ while its conservation implies that $T_{zz} = T_{zz}(z)$ is holomorphic. We define $T(z) = -2\pi T_{zz}$ so that $T_{++} = -T_{zz} = T(z)/(2\pi)$ and
\begin{align}
    \L_{2n} = \int_\mathbb{R}\d x^+\,(x^+)^{2n}T_{++}(x^+) = \frac{(-1)^n}{2\pi i}\int_{i\mathbb{R}}\d z\,z^{2n}T(z)\ .
\end{align}
We can now make a Möbius transformation $w = -(z+1)/(z-1)$ which sends the imaginary axis to the (positively oriented) unit circle $\abs{w}^2 = 1$ and allows us to express $\L_{2n}$ as
\begin{align}
    \L_{2n} = \frac{(-1)^n}{2\pi i}\oint_{\abs{w} = 1}\d w\,\diff{z}{w}z^{2n}\roha{\diff{w}{z}}^2T'(w) = \frac{(-1)^n}{4\pi i}\oint_{\abs{w} = 1}\d w\,\frac{(w-1)^{2n}}{(w+1)^{2n-2}}T'(w)\ , \label{LO_IntegralLO}
\end{align}
where $T' = (\d z/\d w)^2T$ is the energy-momentum tensor in the $w$-coordinate. Closing the contour is only unambiguous for two integer values of $n$, namely $n=0$ or $n=1$; for higher $n$, the contour crosses a pole at $w=-1$. Using \eqref{LO_IntegralLO}, the ANEC operator $\L_0$ can be expressed as
\begin{align}
    \L_0 = \frac{1}{2}\oint_{\abs{w} = 1}\frac{\d w}{2\pi i}(w + 1)^2T'(w) = L_0 + \frac{1}{2}(L_1 + L_{-1})\ , \label{LO_VirasoroANEC}
\end{align}
where the Virasoro operators $L_j$ are defined as $L_j = (2\pi i)^{-1}\oint\d w\,w^{j+1}T'(w)$. We can give a very similar expression for the first even light-ray operator $\L_2$:
\begin{align}
    \L_2 = -\frac{1}{2}\oint_{\abs{w} = 1}\frac{\d w}{2\pi i}\,(w-1)^2T'(w) = L_0 - \frac{1}{2}(L_1 + L_{-1}) \ . \label{LO_Virasorox2ANEC}
\end{align}
Expressing $\L_{2n}$ with $n\geq2$ is more complicated, primarily because the contour crosses a pole at $w=-1$ for these $n$. There are two ways to confront this issue: we deform the contour to include this pole, or we deform it to exclude it (equivalently, one can use an $i\epsilon$ prescription to move the pole into or out of the contour). Since a Möbius transformation can map the interior of a circle to its exterior and vice versa, these approaches should be equivalent, i.e. we should have $T'(w=-1) = 0$ in any correlator if we include the pole in the contour. This is not unreasonable since $w = -1$ corresponds to $x^+\rightarrow\pm\infty$, so $T'(-1) = 0$ expresses the idea that fields should decay sufficiently fast as we approach the conformal boundary of spacetime. 

Assuming that we deform the contour to exclude the pole at $w = -1$, the light-ray operators with $n\geq2$ can be written as
\begin{align}
    \L_{2n} = \frac{(-1)^n}{2}\oint_{\abs{w} = 1}\frac{\d w}{2\pi i}\,\frac{(w-1)^{2n}}{(w+1)^{2n-2}}T'(w) = \sum_{k=1}^{2n}\sum_{l=0}^\infty(-1)^k\roha{\begin{array}{c}
        2n \\
        k
    \end{array}}\roha{\begin{array}{c}
        2-2n \\
        l
    \end{array}}L_{k+l-1}\ ,
\end{align}
where we have Taylor expanded $(w+1)^{-2n+2}$, and a binomial coefficient with a negative upper entry should be interpreted in terms of gamma functions. For simplicity and since the positivity of $\L_0$ is well-established, we will restrict our attention to $\L_2$.

\subsection{Proofs that \texorpdfstring{$\int (x^+)^2 T_{++}\d x^+ \geq 0$}{L2>0} in two-dimensional CFT}\label{subsec_2dCFTproof}
We now consider the second simplest light-ray operator in two-dimensional CFT, $\L_2$. There is a very quick argument that this operator is positive (see e.g. \cite{LO_Mathys}). The ANEC operator $\L_0$ is known to be positive, and an inversion $y = 1/x$ maps $\L_2$ to the ANEC operator:
\begin{equation}
\L_2 = \int_\mathbb{R}\d x^+ (x^+)^2 T_{x^+ x^+} = \int_\mathbb{R} \d y^+\, T_{y^+ y^+} = \L_0\ .
\end{equation}
Inversion is part of the global conformal group, so there is no correction to the classical transformation of the stress tensor. The only potential subtlety with this argument relates to the points $x^+ = 0$ and $x^+\rightarrow\pm\infty$. The region of integration $x^+ \in (-\infty,0)\cup(0,\infty)$ maps to the identical region of $y$ integration, but $x^+ = \pm \infty$ map to $y^+ = 0$ and vice versa. So one might worry that states with interesting behaviour near these special points could possibly lead to a subtlety in the equivalence between $\L_0$ and $\L_2$.

There is a second simple argument that $\L_2$ is positive, suggested to us by Jan de Boer. As shown in \eqref{LO_VirasoroANEC} and \eqref{LO_Virasorox2ANEC}, $\L_0$ and $\L_2$ can be expressed very similarly in terms of the Virasoro operators $L_0$, $L_{-1}$, and $L_1$. Since the ANEC (i.e. the positivity of $\L_0$) can be proven algebraically by writing an arbitrary state as a superposition of primaries and descendants, it is tempting to perform a similar calculation for $\L_2$.

It is most convenient to use a representation of the global conformal group rather than the larger Virasoro algebra, and write a general state as
\begin{equation}
    \ket{\psi} = \sum_{\Delta}\sum_{n=0}^\infty c_{\Delta n} L_{-1}^n \ket{\Delta}\ , \label{LO_L2State}
\end{equation}
where the $c_{\Delta n}$ are free complex constants (chosen to normalise $\ket{\psi}$). We aim to calculate
\begin{equation}
    \mean{\L_2}_{\psi} = \frac{1}{2}\sum_{\Delta,\Delta'} \sum_{n,n'=0}^\infty c^*_{\Delta' n'} c_{\Delta n} \bra{\Delta'} L_{1}^{n'} (2 L_0 - L_1 - L_{-1})  L_{-1}^{n}\ket{\Delta}\ .
\end{equation}
Different primaries do not mix in this transition amplitude, so we can set $\Delta = \Delta'$. The only non-zero terms are 
\begin{equation}
    \mean{\L_2}_{\psi} = \frac{1}{2}\sum_{\Delta}\sum_{n=0}^\infty \bra{\Delta} (2 L_1^n L_0 L_{-1}^n c^*_{\Delta n} c_{\Delta n} - L_1^{n+1} L_{-1} L_{-1}^n c^*_{\Delta n+1} c_{\Delta n} - L_1^{n} L_{1} L_{-1}^{n+1} c^*_{\Delta n}c_{\Delta n+1}) \ket{\Delta}\ .
    \label{sumdn}
\end{equation}
Since the different primaries do not mix, we see that $\L_2$ is positive if and only if its expectation value is non-negative in any state which is a superposition of descendants of a single primary. Let us therefore fix $\Delta$ and suppress the label indicating the primary in the $c$-coefficients, giving
\begin{equation}
    \mean{\L_2}_\psi = \frac{1}{2}\sum_{n=0}^\infty \bra{\Delta} 2 c^*_n c_n L_1^n L_0 L_{-1}^n   - c^*_{n+1} c_{n} L_{-1}^{n+1} L_1^{n+1} - c^*_{n} c_{n+1} L_{1}^{n+1} L_{-1}^{n+1} \ket{\Delta}\ .
\end{equation}
Let us also redefine the $c_n$ to absorb the normalization,
\begin{equation}
    c_n \sqrt{\bra{\Delta} L_1^n L_{-1}^n \ket{\Delta} } \equiv b_n\ ,
\end{equation}
and use that
\begin{equation}
    L_0 L_{-1}^n \ket{\Delta} = (\Delta + n) L_{-1}^n \ket {\Delta} \ ,
\end{equation} 
so that 
\begin{equation}
\expval{ \L_2}_\psi = \frac{1}{2}\sum_{n=0}^\infty \left[ 2(\Delta + n) |b_n|^2 - (b^*_n b_{n+1} + b^*_{n+1}b_n)
\sqrt{\frac{\bra{\Delta} L_1^{n+1} L_{-1}^{n+1} \ket{\Delta}}{\bra{\Delta} L_1^n L_{-1}^n \ket{\Delta}}}\right] \ .
\end{equation}
Now we use 
\begin{equation}
    \bra{\Delta} L_1^{n+1} L_{-1}^{n+1} \ket{\Delta} = (n+1)(2 \Delta + n) \bra{\Delta} L_1^{n} L_{-1}^{n} \ket{\Delta}\ ,
\end{equation}
which can be derived from the global conformal algebra, so that  
\begin{equation}
\expval{ \L_2}_\psi = \frac{1}{2}\sum_{n=0}^\infty \left[ 2(\Delta + n) |b_n|^2 - (b^*_n b_{n+1} + b^*_{n+1}b_n) \sqrt{(n+1) (2 \Delta + n)} \right]
 \ .
\end{equation}
To see that $\L_2$ is positive, we find it helpful to rewrite this as a matrix equation: collecting the $b_n$ into a vector $\vec{b}$, we have
\begin{equation}
  \expval{ \L_2}_\psi = \vec{b}^\dagger M \vec{b}  
\end{equation}
with
\begin{equation}
    M = \begin{pmatrix}
        2 \Delta & -\sqrt{2 \Delta} & 0 & 0 & \dots \\
        -\sqrt{2 \Delta} & 2(\Delta + 1) & -\sqrt{2(2 \Delta + 1)} & 0 & \dots \\
        0 & -\sqrt{2(2 \Delta + 1)} & 2 (\Delta + 2) & -\sqrt{3(2 \Delta + 2)} & \dots \\
        0 & 0 & -\sqrt{3(2 \Delta + 2)} & 2(\Delta + 3) & \dots \\
        \vdots & \vdots & \vdots & \vdots & \ddots
    \end{pmatrix}\ .
\end{equation}
Showing that $\L_2$ is positive is equivalent to showing that $M$ is a positive matrix, in that all eigenvalues are non-negative. This can be accomplished by writing
\begin{equation}
    M = S^\dagger S \ \ \ \ {\rm with} \ \ \ S = \begin{pmatrix}
        \sqrt{2 \Delta} & -1 & 0 & 0 & \dots\\
        0 & \sqrt{2\Delta + 1} & -\sqrt{2} & 0 & \dots \\
        0 & 0 & \sqrt{2\Delta + 2} & -\sqrt{3} & \dots \\
        0 & 0 & 0 & \sqrt{2 \Delta + 3} & \dots \\
        \vdots & \vdots & \vdots & \vdots & \ddots
    \end{pmatrix}\ .
\end{equation}
Matrices of the form $M = S^\dagger S$ are positive, so this establishes that the expectation value is non-negative in all states of the form \eqref{LO_L2State}. Note that a nearly identical argument works for the ANEC operator, which differs from $\L_2$ only by the sign of the off-diagonal terms in \eqref{sumdn}.

A faster argument is as follows: $\L_2$ is related to the ANEC operator $\L_0$ by the replacements $L_{\pm1} \to -L_{\pm1}$. This change of sign leaves the global conformal algebra (which is the relevant algebra as we consider global descendants of quasi-primary states) invariant, and hence $\L_2$ must be positive if the ANEC holds, which we assume\footnote{We thank Jan de Boer for this elegant argument.}. This argument shows that as long as the mapping from the plane to the cylinder does not have any subtleties, $\L_2$ must be positive.

\subsection{(Counter-)example of the positivity of \texorpdfstring{$\L_2$}{L2}} \label{subsec_CFTcounter}

To see how the arguments from the previous section play out in more explicit detail, let us consider the expectation value of $\L_2$ in several concrete states. We find that it is possible to construct a reasonable-looking state in any two-dimensional CFT which violates the positivity of $\L_2$, although we are not able to violate the ANEC.

A commonly considered state is the one that is obtained by acting with a primary operator $\calO_{h,\bar{h}}$, where $(h,\bar{h})$ denotes its conformal weights, at negative Euclidean time $-it$ ($t\geq0$):
\begin{align}
    \ket{\psi} = \int_0^\infty\d t\,f(t)\calO_{h,\bar{h}}(-it,0)\ket{0}\ , \label{LO_ScalarState}
\end{align}
with $f$ a state preparation function; preparing states with Euclidean time evolution combined with operator insertions is standard in CFT. The expectation value of the stress tensor follows from the $T\calO\calO$ three-point function, which is universal and leads to
\begin{align}
    \braket{\psi}\smallmean{T_{++}(x^+)}_\psi \alis \frac{h}{2\pi} \int_0^\infty\frac{\d t\,\d t'\,f^*(t)f(t')}{(t+t')^{2(h+\bar{h} - 1)}(x^+-it)^2(x^+ + it')^2} \label{LO_ScalarStateEMT} \\
    \braket{\psi} \alis \int_0^\infty\frac{\d t\,\d t'\,f^*(t)f(t')}{(t + t')^{2(h+\bar{h})}}\ ,
\end{align}
where we remind the reader that $z = ix^+$. \eqref{LO_ScalarStateEMT} is in general quite cumbersome to compute, so let us instead focus on a particular $f$, e.g. $f(t) = \alpha(2t_0)^{h+\bar{h}}\delta(t - t_0)$ with $t_0 > 0$ and $\alpha\inC$ constants. This choice leads to
\begin{align}
    \smallmean{T_{++}(x^+)}_\psi \alis \frac{2ht_0^2}{\pi\roha{(x^+)^2 + t_0^2}^2}\ ,
\end{align}
which is manifestly positive and decays as $(x^+)^{-4}$ as $x^+\rightarrow \pm \infty$. Thus, $\mean{\L_2}_\psi$ is finite and positive given \eqref{LO_ScalarState} with $f(t) = (2t_0)^{h+\bar{h}}\delta(t - t_0)$.

Another commonly considered state is obtained by inserting the energy-momentum tensor (of which every two-dimensional CFT has one) at negative Euclidean time. In particular, we consider its superposition with the vacuum (we also could have taken a superposition of \eqref{LO_ScalarState} with the vacuum, but since the $T\calO$ two-point function vanishes, \eqref{LO_ScalarStateEMT} would remain unchanged):
\begin{equation}
    \ket{\psi} = \ket{0} + \frac{2\pi}{\sqrt{c}}\int_0^\infty \d t\,f(t)T_{++}(-it, 0) \ket{0}\ , \label{LO_CFTcounterState}
\end{equation}
where $c$ is the central charge and $f$, as before, is the state preparation function. We now have
\begin{align}
    \braket{\psi}\smallmean{T_{++}(x^+)}_\psi \alis \frac{2\pi}{\sqrt{c}}\int_0^\infty\d t\, f(t)\smallmean{T_{++}(x^+) T_{++}(-it)} + \mathrm{c.c.} \non
    &\qquad \qquad \quad + \frac{4\pi^2}{c}\int_0^\infty \d t\, \d t'\, f^*(t')f(t)\smallmean{T_{++}(it') \tpp(x^+) \tpp(-it)} \\
    \expval{\psi | \psi} \alis 1 + \frac{4\pi^2}{c}\int_0^\infty \d t\,\d t'\, f^*(t') f(t) \expval{\tpp(it') \tpp(-it)}\ ,
\end{align}
where the quantities on the right side are all vacuum expectation values. Using the universal two- and three-point functions of the energy-momentum tensor in CFT, this becomes
 \begin{align}
    \braket{\psi}\expval{T_{++}(x_+)}_\psi \alis \frac{\sqrt{c}}{4\pi}\int_0^\infty \frac{\d t\,f(t)}{(x^+ + i t)^4} + \mathrm{c.c.} + \frac{1}{2\pi}\int_0^\infty \frac{\d t\,\d t'\, f^*(t') f(t)}{(t + t')^2 (x^+ + i t)^2 (x^+ - it')^2 } \label{LO_EMTstateEMT}\\
    \expval{\psi | \psi} \alis 1 + \frac{1}{2}\int_0^\infty \frac{\d t\,\d t'\, f^*(t')f(t)}{(t + t')^4}\ ,
\end{align}
This formula depends on the CFT under consideration only through the central charge $c$. As before, we could consider $f(t) = \alpha(2t_0)^2\delta(t-t_0)$ with $t_0 > 0$ and $\alpha\inC$ constants; using the residue theorem, we find that the terms in \eqref{LO_EMTstateEMT} that are linear in $f$ do not contribute to $\mean{\L_2}_\psi$ in this case, leaving only the positive contribution of the final term of \eqref{LO_EMTstateEMT}. 

However, it is clear that the sign of the first terms in \eqref{LO_EMTstateEMT} can be tuned by choosing the phase of $f$ appropriately. We can therefore choose a simple state preparation function to find a counterexample to the positivity of $\L_2$, e.g.
\begin{equation}
    f(t) =  \alpha \Theta(t - t_0)\ , \label{LO_fChoice}
\end{equation}
with $t_0 > 0$ and $\alpha\inC$ constants. Here the $\Theta$-function is taken to be 1 for positive argument and 0 for negative argument. We discuss whether this state should be allowed in section \ref{sec_DiscConc}.

Evaluating the integrals, we have
\begin{align}
    \braket{\psi}\smallmean{T_{++}(x^+)}_\psi \alis \frac{\alpha \sqrt{c} }{12\pi(t_0 - i x^+)^3} + \mathrm{c.c.} + \frac{|\alpha|^2}{2\pi}  \int^\infty_{t_0}\frac{\d t\,\d t'}{(t + t')^2 (x^+ + it)^2 (x^+ - i t')^2 } \\
     \expval{\psi | \psi} \alis 1 + \frac{|\alpha|^2}{2}\int_{t_0}^\infty\frac{\d t\,\d t'}{(t + t')^4} \ .  
\end{align}
The $\alpha^2$-term in the first line is inconvenient to integrate, so for simplicity we work at small $\alpha$. We still need to check that this term gives a finite contribution to $\L_2$. The contribution to $\braket{\psi}\mean{\L_2}_\psi$ of this term is
\begin{equation}
  \frac{|\alpha|^2}{2\pi}\int_{-\infty}^\infty \d x^+ (x^+)^2 \int^\infty_{t_0} \frac{\d t\,\d t'}{(t + t')^2 (x^+ + it)^2 (x^+ - i t')^2 } \ .
\end{equation}
The integral converges absolutely, so we can interchange the order of integration and do the $x^+$ integral by contours, yielding 
\begin{equation}
    \frac{|\alpha|^2}{2\pi} \int_{-\infty}^\infty \d x^+ (x^+)^2  \int^\infty_{t_0} \frac{\d t\,\d t'}{(t + t')^2 (x^+ + it)^2 (x^+ - i t')^2 }  \sim |\alpha|^2 \int_{t_0}^\infty \frac{t t'\, \d t \d t'}{(t+ t')^5} \ .
\end{equation}
The remaining integral is finite:
\begin{equation}
    \int_{t_0}^\infty \frac{t t'\, \d t \d t'}{(t+ t')^5} \sim \frac{1}{t_0}\ .
\end{equation}
Similarly, we find that the $\alpha^2$-term in the norm of the state is finite. Having established this, we work to first order in $\alpha$, finding
\begin{equation}
    \braket{\psi}\expval{T_{++}(x_+)}_\psi = \frac{\alpha\sqrt{c}}{12\pi(t_0 - i x^+)^3} + \mathrm{c.c.} + \order{\alpha^2}\ ;
\end{equation}
this allows us to evaluate the expectation value of $\L_2$, namely
\begin{equation}
    \braket{\psi}\expval{\L_2}_\psi = \frac{\alpha\sqrt{c}}{12\pi}\int_{-\infty}^\infty \frac{\d x^+\,(x^+)^2}{(t_0 - i x^+)^3} + \mathrm{c.c.} + \order{\alpha^2}\ . \label{LO_IntegralL2}
\end{equation}
This integral diverges due to the large-$x^+$ behaviour of the integrand unless we choose $\alpha\inR$; then, the integral converges and we obtain
\begin{equation}
    \braket{\psi}\expval{\L_2 }_\psi = -\frac{\alpha\sqrt{c}}{6} + \order{\alpha^2} \ .
\end{equation}
Since this term is linear in $\alpha$, it can be chosen to have either sign. Therefore, this simple state gives a negative expectation value for the $\L_2$ operator.

\section{Discussion and conclusions}\label{sec_DiscConc}

In the preceding sections, we have given both proofs and counterexamples for the hypothesis that the light-ray operators $\L_{2n}$ from \eqref{LO_def} are positive operators, in the context of a non-minimally coupled free scalar field and general two-dimensional CFTs. The counterexamples served to illustrate the properties that characterise the set of states for which $\L_{2n}$ is positive. In this section, we will discuss the relationship between the various proofs and their counterexamples in more detail, as well as some implications of the positivity of the light-ray operators.

\subsection{The free scalar field}
The most important assumption about the class of states to which the proof in section \ref{subsec_FreeProof} applies, is that modes with vanishing or divergent lightcone momentum $p_+$ are sufficiently suppressed. This assumption allowed us to discard the off-diagonal term in \eqref{LO_Volledig} and subsequently ignore any boundary terms when integrating by parts. We also argued that it is reasonable: for massive fields, it corresponds to the assumption that modes with infinite momentum are sufficiently suppressed. For massless fields in $d\geq3$ spacetime dimensions, it is equivalent to the statement that expectation values of creation and annihilation operators should be sufficiently smooth (implying that no mode has a divergent occupation number) as well as sufficiently suppressed at high momenta.

Despite this apparently reasonable condition, section \ref{subsec_Counterexample} identified a class of states which satisfy several acceptability criteria (as Hadamard states with finite energy and norm) but still yield an indefinite expectation value of $\L_{2n}$. As shown at the end of section \ref{subsec_Counterexample}, this happens because these states have insufficiently suppressed support at vanishing lightcone momentum: as $p_+ + k_+ \rightarrow 0$, $h$ from \eqref{LO_Counter_h} vanishes polynomially, so one of its derivatives does not vanish at $p_+ = k_+ = 0$. To further emphasise this point, note from \eqref{LO_LOCross} that choosing $a$ and $b$ such that $\mean{\L_{2n}}_{\psi}$ is indefinite prevents $T_{++}^\mathrm{cross}$ from contributing to $\mean{\L_{2m}}_{\psi}$ if $m<n$. Instead, $\mean{\L_{2m}}_\psi$ is determined by $T_{++}^\mathrm{diag}$ with a sign that cannot be tuned, recovering the result of section \ref{subsec_FreeProof}. The reason for this is suggested by \eqref{LO_CrosstermDerivative}: lower light-ray operators are less demanding with regard to the rate at which modes of vanishing lightcone momentum should be suppressed.

On the other hand, the contribution of $T_{++}^\mathrm{cross}$ to $\mean{\L_{2M}}_{\psi}$ with $M>n$ is divergent (as can be seen in \eqref{LO_LOCross} by expanding $(x^+)^{2M+2}$ around $il$), meaning that $\L_{2M}$ is not well-defined as an operator in this particular state. Combining this fact with our observations on $\L_{2m}$ with $m<n$, it seems that the counterexamples from section \ref{subsec_Counterexample} are quite special: states which are counterexamples to the positivity of $\L_{2n}$ are not counterexamples to the positivity of $\L_{2m}$ with $m < n$, while $\mean{\L_{2M}}_\psi$ with $M > n$ diverges in these states. If this is not an artefact of our class of states and our specific theory, then we could give an alternative characterisation of the combination of states and light-ray operators for which the computation in section \ref{subsec_FreeProof} holds: namely, that $\mean{\L_{2n}}_{\psi}\geq0$ when $n$ satisfies \eqref{LO_NonMinConstraint} and $\abs{\smallmean{\L_{2(n+1)}}_{\psi}} < \infty$.

Furthermore, let us mention two more physical ramifications of the proof and its counterexamples in section \ref{sec_FreeField}. First, one of our motivations to study light-ray operators was that the CANEC, applied to the lightcone in Minkowski spacetime, would imply that $\L_d$ is positive for holographic CFTs for even $d\geq4$ \cite{CANEC_odd,CANEC_even} (for $d=2$, \cite{CANEC_even} reproduces the ANEC). However, \ref{subsec_Counterexample} provides counterexamples to such a statement for the conformally coupled scalar field. This implies at least one of three things: the strong coupling which characterises holographic CFTs renders $\L_d$ positive, the CANEC only applies to a restricted set of states, or the CANEC does not apply to all congruences in all spacetime geometries. Since \cite{CANEC_odd,CANEC_even} only gave a holographic proof of the CANEC in a limited class of geometries, it stands to reason that at least the latter possibility is true; assessing the former two will require further investigation.

Our second point is concerned with the Rindler vacuum. This state has a negative null energy density everywhere except at the origin, and thus forms another counterexample to the positivity of $\L_{2n}$; this implies that it cannot satisfy the assumptions we made in section \ref{subsec_FreeProof}. This raises the question of how one should think about the states for which $\L_{2n}$ is indefinite: are they unphysical, or are they still relevant to the theory? In the former case, it would be necessary to explain why an observer accelerating at a constant rate could not access their natural vacuum, and in the latter case one would need to explain what (if anything) physically distinguishes states for which $\L_{2n}$ is positive from those with indefinite $\L_{2n}$.  

A way to circumvent the problem posed by the Rindler vacuum would be to argue that it is not realistic due to its divergent energy density at a single point. However, we could consider a state which approximates the Rindles vacuum away from the origin and behaves more smoothly near it. Such a smoothed Rindler vacuum could still be a counterexample to the positivity of $\L_{2n}$, since $\L_{2n}$ suppresses the (positive) contributions from the region near the origin. On the other hand, smoothing out the divergence near $x^+ = 0$ might change the null energy density at large $\abs{x^+}$ in such a way that $\L_{2n}$ stays positive. In any case, the physical interpretation of states with indefinite $\L_{2n}$ remains an interesting avenue of further research.

\subsection{ \texorpdfstring{$d=2$}{d=2} conformal field theory}

Having concluded our treatment of the free, non-minimally coupled scalar field, we went on to consider light-ray operators in a $d=2$ CFT in section \ref{sec_CFT}. This is an important complement to the free theories from section \ref{sec_FreeField}, since they correspond to massless fields in two dimensions (the only context in which the free field computation does not apply) and allow for interactions. We focused our efforts on $\L_2$, since it is closely related to the ANEC operator and, as shown in \eqref{LO_Virasorox2ANEC}, it has a very simple expression in terms of the Virasoro operators.

We then proved the positivity of $\L_2$ in three ways. First, we mapped it to the ANEC operator with an inversion, which assumed that expectation values of $T_{++}$ vanish sufficiently fast as $x^+\rightarrow\pm\infty$; this is similar to our assumptions for the free field. The other proofs expressed $\L_2$ in terms of Virasoro operators, which are well-defined in the theory on the (Lorentzian) cylinder or in radial quantisation on the plane. Positivity of $\L_2$ in the representation of the global conformal group then followed by explicitly calculating its expectation value and by relating it to the ANEC operator via a change of sign for $L_{\pm1}$. 

Nevertheless, in section \ref{subsec_CFTcounter} we have found a set of states that allow for negative expectation values of $\L_2$, meaning that they must somehow violate the assumptions of our proofs. This is most straightforwardly interpreted for our algebraic proofs, which assumed that we work with states that are well-defined on the cylinder (i.e. states from the representation of the global conformal group). Apparently, the state \eqref{LO_CFTcounterState} with \eqref{LO_fChoice} is not a good state on the cylinder.

However, from the point of view of the theory on the plane quantized on equal-time surfaces (e.g. $t=0$) rather than in radial quantization, it is not clear that there is any problem with this state. The state is normalizable and has finite energy. Because it is prepared by evolving in Euclidean time, with operator insertions away from zero Euclidean time, the short distance behaviour is well-behaved, meaning that the state satisfies the Hadamard condition. From the point of view of the theory on the plane, there does not seem to be any justification for prohibiting this class of states.

A natural question, then, is whether there is a simple definition of the class of states for which $\L_2$ is positive. We can give some data points with regards to this question. Consider again our state $\ket{\psi}$ defined in \eqref{LO_CFTcounterState}. In this state, we have
\begin{align}
    \braket{\psi} \expval{\L_2}_\psi \alis \int_{-\infty}^\infty\!\frac{\d x^+ (x^+)^2}{2\pi}\! \left[\frac{\sqrt{c}}{2}\!\int_0^\infty\!\frac{\d t\,f(t)}{(x^+ + i t)^4} + \mathrm{c.c.} +\! \int_0^\infty\!\alpha\d\alpha\,\abs{\int_0^\infty\!\frac{\d t\,f(t)e^{-\alpha t}}{(x^+ + it)^2}}^2\right] , \label{LO_CFTL2General}
\end{align}
where we applied \eqref{LO_GammaIdentity} to the last term of \eqref{LO_EMTstateEMT} to make its positivity manifest. In fact, observe that the $t$-integral in this last term is the Laplace transform of $f(t)/(x^+ + it)^2$; by Lerch's theorem \cite{InverseLaplace} this vanishes if and only if $f$ is a null function, i.e. its integral vanishes on any subset of its domain. Thus, negative contributions to $\mean{\L_2}_\psi$ from the terms linear in the state preparation function $f$ are always at least partially compensated by a positive contribution from the last term in \eqref{LO_CFTL2General}. Now, suppose that $f$ is sufficiently well-behaved that the order of integration can be interchanged in the terms linear in $f$. In that case, we can perform the $x^+$-integral first, obtaining
\begin{equation}
    \int_{-\infty}^\infty\frac{\d x^+\,(x^+)^2}{(x^+ + i t)^4} = 0\ ;
\end{equation}
so as long as the order of integration can be interchanged, the terms linear in $f$ give zero contribution, and $\L_2$ is positive. The condition for changing the order of integration is given by the Fubini theorem, which states that we can change the order of integration if
\begin{equation}
    \int_{-\infty}^\infty\!\d x^+\int_0^\infty\!\d t\ \abs{\frac{(x^+)^2 f(t)}{(x^+ + i t)^4}} < \infty\ . \label{LO_Fubini}
\end{equation}
This will be the case as long as $f(t)$ vanishes sufficiently fast as $t \to 0$ and as $t \to \infty$. The state preparation function \eqref{LO_fChoice} violates the latter condition, and indeed, it provides a simple counterexample to the positivity of $\L_2$.

Extrapolating from these data points, one might guess that $\L_2$ is positive for any state prepared by acting with operators that are compactly supported away from $t=0$ in the lower half Euclidean plane. This is consistent with the proof for the positivity of $\L_2$: any such state is well-defined on the cylinder and therefore satisfies the assumptions of section \ref{subsec_2dCFTproof}. Additionally, we find that the state created by a single operator insertion in \eqref{LO_ScalarState} has $\mean{\L_2}_\psi\geq0$ as well if $f$ is compactly supported away from $t=0$. After all, if $f$ is compactly supported, \eqref{LO_ScalarStateEMT} shows that $\mean{T_{++}}_\psi$ decays as $(x^+)^{-4}$ for $x^+\rightarrow\pm\infty$; if $f$ is not supported at $t=0$, this in turn implies that the $x^+$- and $t$-integrals involved in $\mean{\L_2}_\psi$ converge absolutely and can be exchanged. The $x^+$-integral can then be evaluated by closing the contour around e.g. the pole at $x^+ = it$ (the decay as $x^+\rightarrow\infty$ is fast enough to allow this), leading to
\begin{align}
    \braket{\psi}\mean{\L_2}_\psi \alis 2h\int_0^\infty\frac{\d t\,\d t'\,f^*(t)f(t')t't}{(t + t')^{2(h + \bar{h}) + 1}} = 2h\int_0^\infty \frac{\d\alpha\,\alpha^{2(h + \bar{h})}}{\Gamma(2h + 2\bar{h} + 1)}\abs{\int_0^\infty\d t\,f(t)e^{-\alpha t}t}^2\ , \label{LO_ScalarL2}
\end{align}
which is manifestly positive. As before, Lerch's theorem implies that the $t$-integral, which is the Laplace transform of $f(t)t$, vanishes if and only if $f(t)t$ is at most a null function. In practice, this means that \eqref{LO_ScalarL2} only saturates $\mean{\L_2}_\psi\geq0$ as \eqref{LO_ScalarState} approaches the vacuum.

Having studied $\L_2$ in some detail, we also offer some comments on the other light-ray operators $\L_{2n}$ in two-dimensional CFTs. Consider for example $\mean{\L_{2n}}_\psi$ for $n\geq2$ and the counterexample state $\ket{\psi}$ from section \ref{subsec_CFTcounter} (i.e. \eqref{LO_CFTcounterState} with \eqref{LO_fChoice}); from \eqref{LO_IntegralL2} one can immediately read off that even at first order in $\alpha$, $\mean{\L_{2n}}_\psi$ diverges for $n\geq2$. This is consistent with the pattern we noted for the counterexamples to the positivity of $\L_{2n}$ in free field theory: for $\ket{\psi}$ from \eqref{LO_CFTcounterState}, $\L_0$ is positive, $\L_2$ is indefinite but finite, and $\L_{2n}$ with $n\geq2$ is divergent.

This divergence is not unique to the $\ket{\psi}$ from \eqref{LO_CFTcounterState}: because light-ray operators involve improper integrals with positive powers of $x^+$, we expect there to be many seemingly sensible states where the expectation value of the light-ray operator diverges. A simple example is \eqref{LO_ScalarState}, a state created with a single compactly supported primary operator insertion; as noted before, \eqref{LO_ScalarStateEMT} then implies that $\mean{T_{++}}_\psi \sim (x^+)^{-4}$ asymptotically, which means that $\L_0$ and $\L_2$ are the only finite light-ray operators in this state.

One way to counteract this divergence is to combine insertions of $\calO_{h,\bar{h}}$ with insertions of the energy-momentum tensor. For example, to construct a state in which $\mean{\L_4}_\psi$ is finite we could consider a superposition of \eqref{LO_ScalarState} and \eqref{LO_CFTcounterState}, choosing the state preparation functions proportional to delta functions:
\begin{align}
    \ket{\psi} = \ket{0} + \frac{2\pi\alpha_1(2t_1)^2}{\sqrt{c}}T_{++}(-it_1,0)\ket{0} + \alpha_2(2t_2)^{h+\bar{h}}\calO_{h,\bar{h}}(-it_2,0)\ket{0} \label{LO_CounterL4State}
\end{align}
with $\alpha_1,\alpha_2\inC$ and $t_1,t_2>0$ constants. Because the $T\calO$- and $TT\calO$-correlators vanish, the expectation value $\mean{T_{++}(x^+)}_\psi$ is straightforwardly computed:
\begin{align}
    \braket{\psi}\smallmean{T_{++}(x^+)}_\psi = \frac{\alpha_1t_1^2\sqrt{c}}{\pi(x^+ + it_1)^4} + \mathrm{c.c.} + \frac{2\abs{\alpha_1}^2t_1^2}{\pi\abs{x^+ + it_1}^4} + \frac{2h\abs{\alpha_2}^2t_2^2}{\pi\abs{x^+ + it_2}^4}\ . \label{LO_CounterL4emt}
\end{align}
To have a finite $\mean{\L_4}_\psi$, we should now demand that $\mean{T_{++}}_\psi$ asymptotically decays faster than $(x^+)^{-5}$. This condition requires $\alpha_1\inR$, with either of
\begin{align}
    \alpha_1 = -\frac{1}{2}\sqrt{c}\, \pm\, \sqrt{\frac{c}{4} - h\abs{\alpha_2}^2\roha{\frac{t_2}{t_1}}^2}\ . \label{LO_L4ConvergenceParameter}
\end{align}
With this relation in mind, we can compute $\braket{\psi}\mean{\L_4}_\psi$ explicitly using the residue theorem:
\begin{align}
    \braket{\psi}\mean{\L_4}_\psi =  3h\abs{\alpha_2t_2}^2\roha{t_1 - t_2} - 5\sqrt{c}\,\alpha_1t_1^3\ .
\end{align}
By fixing $\alpha_2t_2$ and considering $t_2\gg t_1$, the negative first term of this expression can be seen to dominate. Therefore, \eqref{LO_CounterL4State} with \eqref{LO_L4ConvergenceParameter} represents a finite counterexample to the positivity of $\L_4$. Note that, given \eqref{LO_CounterL4emt}, this state results in a positive $\mean{\L_2}_\psi$. Again, this is consistent with our observation from free field theory that a finite $\smallmean{\L_{2(n+1)}}$ seemingly implies a positive $\mean{\L_{2n}}$.

Nevertheless, the fact that even an innocuous-looking state like \eqref{LO_ScalarState} leads to divergent $\mean{\L_{2n\geq4}}$ may make it seem like only fine-tuned states have finite expectation values of the light-ray operators. However, consider the set of states we can actually prepare in CFT. We are restricted to act in Lorentzian time with unitary operators, so a natural set of states to consider is the states obtained by acting with sources $j_i$ in Lorentzian time:
\begin{equation}
    \ket{\psi} = \exp(i \int \d^2x \sum_k j_k(x) \mathcal{O}_k (x)) \ket{0}\ . \label{LO_LorentzStates}
\end{equation}
Here, the integral is over Lorentzian spacetime, and we have suppressed tensor indices. Suppose now that we make the natural assumption that we can only turn on the sources in some finite spacetime region, i.e. the $j_k(x)$ are compactly supported. Because $T_{++}$ is holomorphic, we can compute $\expval{T_{++}(x^+)}$ on the line $t=0$. Since operators at large $x^+$ commute with all operators in the state preparation, the Campbell identity implies that the set of states in \eqref{LO_LorentzStates} has
\begin{equation}
    \smallmean{T_{++}(x^+)}_\psi = 0 \label{LO_VanishingEMT}
\end{equation}
for sufficiently large $|x^+|$. Therefore, in the natural set of states \eqref{LO_LorentzStates}, \textit{all} light-ray operators have finite expectation values. If the relation between the finiteness of $\L_{2(n+1)}$ and the positivity of $\L_{2n}$ which we observed in free field theory holds for two-dimensional CFTs as well, then \eqref{LO_VanishingEMT} implies that all light-ray operators are positive in the set of states \eqref{LO_LorentzStates}. We leave the question of whether this is indeed the case to future work.

\section*{Acknowledgements}
We thank Jan de Boer, Diego M. Hofman, Ken D. Olum, David Simmons-Duffin, and Anthony J. Speranza for useful discussions. BF is partially supported by Heising-Simons Foundation ‘Observational Signatures of Quantum Gravity’ QuRIOS collaboration grant.

\bibliographystyle{unsrt}
\bibliography{sources}

\end{document}